\documentclass[journal,comsoc]{IEEEtran}
\usepackage{color}

\usepackage{cite}

\ifCLASSINFOpdf
\else
\fi
\usepackage[cmex10]{amsmath}
\usepackage{amssymb}
\usepackage{mathtools}
\usepackage{amsthm}
\usepackage{amsmath}
\usepackage{graphicx}
\usepackage{bm}
\usepackage{subfigure}
\usepackage{setspace}
\graphicspath{{figures/}}
\newcounter{qcounter}
\begin{document}
\title{Multidimensional Constellations for Uplink SCMA Systems --- A Comparative Study}
\author{Monirosharieh Vameghestahbanati,~
Ian Marsland,~\IEEEmembership{Member,~IEEE,}
Ramy H. Gohary,~\IEEEmembership{Senior Member,~IEEE,}
 and Halim Yanikomeroglu,~\IEEEmembership{Fellow,~IEEE}\thanks{This work is supported in part by an Ontario Trillium Scholarship, and in part by Huawei Canada Co., Ltd.. 

The authors are with the Department of Systems and Computer Engineering, Carleton University, Ottawa, ON K1S 5B6, Canada (email: mvamegh@sce.carleton.ca; ianm@sce.carleton.ca; gohary@sce.carleton.ca; halim@sce.carleton.ca).}}



\maketitle
\begin{abstract}
Sparse code multiple access (SCMA) is a class of non-orthogonal multiple access (NOMA) that is proposed to support uplink machine-type communication services. In an SCMA system, designing multidimensional constellation plays an important role in the performance of the system. Since the behavior of multidimensional constellations highly depends on the type of the channel, it is crucial to employ a constellation that is suitable for a certain application. {\color{black}In this paper, we first highlight and review the key performance indicators (KPIs) of multidimensional constellations that should be considered in their design process for various channel scenarios. We then provide a survey on the known multidimensional constellations in the context of SCMA systems with their design criteria. The performance of some of those constellations are evaluated for uncoded, high-rate, and low-rate LTE turbo-coded SCMA systems under different channel conditions through extensive simulations. } 
 All turbo-coded comparisons are performed for bit-interleaved coded modulation, with a concatenated detection and decoding scheme. Simulation results confirm that multidimensional constellations that satisfy KPIs of a certain channel scenario outperform others. Moreover, the bit error rate performance of uncoded systems, and the performance of the coded systems are coupled to their bit-labeling. The performance of the systems also remarkably depends on the behavior of the multi-user detector at different signal-to-noise ratio regions. 

\end{abstract}

\begin{IEEEkeywords}
Non-orthogonal multiple access (NOMA), Sparse code multiple access (SCMA), low density spreading (LDS), multidimensional constellation, SCMA codebook, fading channels, message passing algorithm (MPA), bit-interleaved coded modulation (BICM), key performance indicators (KPIs). 
\end{IEEEkeywords}

\IEEEpeerreviewmaketitle

\section{Introduction}\label{sec:Intro}
\IEEEPARstart{O}{ne} key difference among the generations of wireless systems from 1G to 4G arises in various multiple access schemes. That is, 1G operates with frequency division multiple access (FDMA), 2G with time division multiple access (TDMA), 3G with code division multiple access (CDMA), and 4G with orthogonal frequency division multiple access (OFDMA). {\color{black}FDMA, TDMA, CDMA, and OFDMA are generally orthogonal multiple access (OMA) schemes, wherein orthogonal resources (time, frequency, or code) are allocated to different users to avoid the interference among them\footnote{{\color{black}To be precise, in 3G, the downlink uses orthogonal CDMA (Walsh-Hadamard codes); but the uplink in general uses non-orthogonal CDMA through pseudo noise-codes (PN-codes). This is due to the fact that orthogonality requires chip-level synchronization; this is next-to-trivial to achieve in the downlink, while very difficult in the uplink. That is, even if the orthogonal Walsh-Hadamard codes are used in the uplink, since the chip-level synchronization is lost due to different propagation times, the received signals at the BS become no longer chip-level synchronized, i.e., the orthogonality is not maintained. As such, other design alternatives that perform better in the absence of chip-level synchronization are used instead of the Walsh-Hadamard codes.}}\cite{Yoan16,Bayestehmagezine18}. However, to provide even higher spectral efficiency, future wireless systems are likely to employ non-orthogonal multiple access (NOMA)\cite{saito13noma,Ding17magazine,shafi175g,shin17magazine}. NOMA allocates resource elements (REs) to users in a non-orthogonal fashion, allowing multiple users to share the same REs. Hence, it improves the spectral efficiency of the system, meeting the needs of the ever-growing demand for mobile Internet and the Internet of Things (IoT)~\cite{Daimagazine, wang17jsac}.}  

{\color{black}The various NOMA techniques that exist by far can be categorized into three main classes \cite{Cai18CST}: multiple-domain NOMA, power-domain NOMA, and code-domain NOMA. 

The power-domain NOMA (also referred as the single-carrier NOMA \cite{Ding17survey}) can support multiple users within the same RE by allocating different power levels to them {\color{black} based on their channel conditions} \cite{choi14,octavia17,guo2017design}. Successive interference cancellation (SIC) is used at the receiver to detect the transmitted signals. It is shown in \cite{chen2016mathematical} that if both NOMA and OMA employ optimal
resource allocation, NOMA can still outperform the conventional OMA. Power-domain NOMA is considered as a potential scheme for 5G systems, specially for downlink scenarios \cite{najafi2017non,Cai18CST}.

Code-domain NOMA, also referred to as multi-carrier NOMA in \cite{Ding17survey}, can serve multiple users within the same RE. In code-domain NOMA, users are differentiated by their different codes.}
Low density signature/spreading (LDS) \cite{hoshyar08,hoshyar10,Imri12,imran2012low,Imari14,razavi11} is one of the early code-domain NOMA techniques, where each user's data is spread over multiple REs in a \emph{sparse} manner. In other words, each user is assigned to only a subset of the available REs. As such, the multi-user interference pattern at the receiver entails a low density graph \cite{Montanari06}, reducing the complexity of the multi-user detection at the receiver. In LDS,  all users employ the repetition of quadrature amplitude modulation (QAM) constellations to spread their information over their pre-assigned REs. {\color{black}At the receiver, due to the sparseness of LDS, the widely used message passing algorithm (MPA) \cite{Kschischang01} can be used for detection.}

Sparse code multiple access (SCMA) \cite{scma} is a novel code-domain NOMA scheme that is essentially an enhanced variation of LDS. {\color{black}Like with LDS}, each RE is shared among only a subset of {\color{black}the} active users, {\color{black} constituting a \emph{sparse} structure}{\color{black}, but} SCMA users map their incoming bits to multidimensional constellations to spread their information {\color{black}instead of using the repetition of QAM over their pre-assigned REs. Therefore, LDS can be considered as a special class of SCMA. }

Multiple-domain NOMA superimposes multiple users in multiple domain, such as the power domain, the code domain, and the spatial domain \cite{Cai18CST}. Pattern division multiple access (PDMA) \cite{pdma16} and lattice partition multiple access (LPMA) \cite{lpma16} are two examples of this type of NOMA.
{\color{black}In PDMA, similar to SCMA, the users spread their data over their pre-assigned REs in a sparse manner. However, unlike SCMA, the number of REs assigned to each PDMA user can vary. Further, PDMA users can be multiplexed in power or space domains as well. When different clusters of users are multiplexed in different domains, MPA followed by SIC is employed for detection \cite{Cai18CST}.
In {\color{black}LPMA}, power and code domains are combined to multiplex users. A multilevel lattice code assigns different lattice codes to different users based on their channel conditions. At the receiver and similar to power-domain NOMA, SIC can be used for detection.

{\color{black}The comparison among different NOMA techniques has been widely studied from different aspects in e.g., \cite{Cai18CST,wang15nomacomparison,moltafet18tvt,wu18nomacomparisonsaccess,Bayestehmagezine18}. It is found in \cite{Cai18CST} that the structure of power-domain NOMA is uncomplicated, and thus, it can be combined with other different technologies like multiple-input multiple output (MIMO) or cooperative networks. On the other hand, to order the users based on their channel conditions, clustering and pairing of users are required, which will add to the complexity of the system. Code-domain NOMA and Multiple-domain NOMA do not require the channel conditions of different users. Moreover, the near-optimal MPA detection used in code-domain and multiple-domain NOMA can perform better than SIC used in power-domain NOMA, but MPA is more complex. In addition, the coding that is being used in code-domain NOMA and some multiple-domain NOMA techniques suggests some redundancies, which reduces the spectral efficiency of the system. In \cite{wang15nomacomparison}, it is shown that compared with PDMA, SCMA achieves a better performance due to its multidimensional constellation. In \cite{moltafet18tvt}, the power-domain NOMA and SCMA are compared from the sum-rate perspective. The numerical analysis in \cite{moltafet18tvt} shows that SCMA achieves a better sum-rate compared with power-domain NOMA, but at the cost of a higher system complexity. A comprehensive study of different NOMA schemes is carried out in \cite{wu18nomacomparisonsaccess}. It is found in \cite{wu18nomacomparisonsaccess} that the code-domain NOMA can achieve a higher robustness and is a better candidate for massive connectivity. In \cite{Bayestehmagezine18}, a systematic overview on the design of different NOMA schemes along with the related standardization process for the next generation of wireless networks are provided. Furthermore, SCMA is presented as a potential uplink NOMA scheme that is proposed in 3GPP Release-14 for the system design of the new radio (NR)\cite{Bayestehmagezine18}.

Due to the attractive features of code-domain NOMA, and particulary SCMA, it is important to investigate SCMA from different aspects.  One of the most appealing applications of SCMA systems is in uplink. In particular, SCMA has been proposed to support uplink machine-type communication services \cite{Yoan16,wf16,Bayestehmagezine18}. As such, the focus of this paper is on uplink SCMA systems. }

As SCMA users map their incoming bits to multidimensional constellations, designing good multidimensional constellations plays an important role in their performance. In fact, it is a key feature that distinguishes SCMA from other NOMA techniques \cite{bayesteh16scmasurvey}.
The design of multidimensional constellation has been studied for downlink SCMA systems in e.g., \cite{Cai16,zhou17,alam17wcnc,alam17icc,Laidownlinkaccess,Yu16starqam,yu17starqamber,Klimentyev17error,Li18}, and for uplink SCMA systems in e.g., \cite{taherzadeh14,Taherzadeh16patent,zhang16orig,Yan16lattice,Yan17access,cqam17,Baoconstellation,Bao16sphericalcode,zhai17,Bao18,He17ldpcscma,nikopour2016systems,peng17,Klimentyevcodebook17,Kulhandjian17}.
The objective of this work is to provide a thorough study on multidimensional constellations for uplink SCMA systems, and to highlight their design metrics under various channel scenarios. The main contributions of the paper can be summarized as follows:
\begin{itemize}
  \item We highlight and review the key performance indicators (KPIs) of multidimensional constellations that should be considered in their design process under various channel scenarios.
  \item We provide an overview of the multidimensional constellations proposed for uplink SCMA systems in \cite{taherzadeh14,Taherzadeh16patent,zhang16orig,cqam17,Baoconstellation,nikopour2016systems,Bao16sphericalcode,Yan16lattice,peng17,Yan17access,Klimentyevcodebook17,zhai17,Kulhandjian17,He17ldpcscma,Bao18} with their design criteria.
  \item The performance of those constellations are evaluated for uncoded, high-rate, and low-rate LTE turbo-coded SCMA systems over different channel conditions through extensive simulations.
 \end{itemize}}
 It is assumed that the channel state information (CSI) is not available at the transmitter. We use the widely used near-optimal message passing algorithm in the logarithmic domain (Log-MPA) \cite{Kschischang01,wei2016low} for the multi-user detection, and {\color{black}perform} the multiuser detection and the turbo decoding separately in a concatenated manner as in \cite{scma,taherzadeh14,vameghestahbanati2017polar}. All turbo-coded comparisons are performed for bit-interleaved coded modulation (BICM). Simulation results confirm that multidimensional constellations that satisfy KPIs of a certain channel scenario outperform other multidimensional constellations in that scenario. Moreover, the bit error rate (BER) performance of uncoded systems, and the performance of the coded systems are tied to their bit-labeling. The performance of the systems also depends on the behavior of the multi-user detector at different signal-to-noise ratio regions. We also present a number of possible research directions.

The rest of this paper is organized as follows. We provide the uplink SCMA system model in Section \ref{sec:sysmodel}{\color{black}, and the SCMA multidimensional design approach in Section \ref{sec:constdesign}}. In Section \ref{sec:chnmodelKPI}, we introduce the channel models of interest, and review the KPIs of multidimensional constellations under those channel scenarios. We then provide an overview {\color{black}of} the multidimensional constellations proposed for SCMA systems in Section \ref{sec:constellations}. In Section \ref{sec:sim}, we evaluate the performance of the system under different conditions. Section \ref{sec:conlusion} concludes the paper, and {\color{black}suggests some} future research directions. 
\section{Uplink SCMA System Model}\label{sec:sysmodel}
\begin{figure*}[!t]
\centering
{\includegraphics[width= 0.78\textwidth]{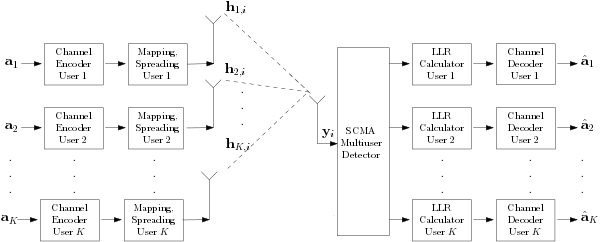}}
\caption{{\color{black}The uplink SCMA system model with $K$ users.}}
\label{fig:scmasys}
\end{figure*}
Consider an uplink SCMA system with $K$ users and $N$ orthogonal resource elements (REs), where $N<K$, and each user is assigned to only $d_v\ll N$ REs out of the $N$ REs. {\color{black}In other words, each user spreads its data over $d_v$ REs.} {\color{black}Under the constraint that no two users should be assigned all the same REs, a system is fully loaded if $K = {N \choose d_v}$. 

The allocation of the $d_v$ REs to {\color{black}the $k^{\textrm{th}}$} user is performed through the $N\times d_v$ binary user-to-RE mapping matrix, $\mathbf{F}_k$. As an example, for a fully-loaded system with $K=6$, $N=4$ and $d_v=2$, the user-to-RE mapping matrix of the $k^{\textrm{th}}$ user, can be
\begin{eqnarray}\label{fk}
\mathbf{F}_k = \left[ {\begin{array}{*{20}c}
   0 & 0 \\
   1 & 0 \\
   0 & 0 \\
   0 & 1 \\
\end{array}} \right].
\end{eqnarray}
The $N$-dimensional binary user-to-RE indicator column vector for each user is then defined by the column vector {\color{black}$\mathbf{s}_k = \textrm{diag}~(\mathbf{F}_k\mathbf{F}_k^{\textrm{T}})$}, where a value of 1 in the $n^\textrm{th}$ elements of $\mathbf{s}_k$ indicates that the $n^{\textrm{th}}$ RE has been assigned to the $k^{\textrm{th}}$ user. In the previous example, $\mathbf{s}_k = [0~1~0~1]^{\textrm{T}}$. That is, user $k$ is assigned to only two REs out of the four available REs, namely, the second and the fourth REs.  
 The user-to-RE indicator matrix, $\mathbf{S}=[\mathbf{s}_1\dots \mathbf{s}_K]$ constitutes an $N\times K$ matrix with each of its columns corresponding to each user with $d_v$ non-zero elements, identifying the $d_v$ REs that are assigned to each user. In a similar vein, each row of $\mathbf{S}$ corresponds to each RE with $d_c$ non-zero elements, identifying the users connected to each RE. {\color{black}By design, $d_v\ll N$ so} $\mathbf{S}$ is a sparse matrix. {\color{black}For instance, consider the user-to-RE indicator matrix of an SCMA system with $K = 6$, $N = 4$, and $d_v=2$:
\begin{eqnarray}\label{S}
\mathbf{S} = \left[ {\begin{array}{*{20}c}
   0 & 1 & 1 & 0 & 1 & 0  \\
   1 & 0 & 1 & 0 & 0 & 1  \\
   0 & 1 & 0 & 1 & 0 & 1  \\
   1 & 0 & 0 & 1 & 1 & 0  \\
\end{array}} \right].
\end{eqnarray}
The first column of $\mathbf{S}$ corresponds to the first user, which is assigned to the second and the fourth REs; the second column of $\mathbf{S}$ corresponds to the second user that is connected to the first and the third REs, and so on.}} {\color{black}Furthermore, the first row of $\mathbf{S}$ corresponds to the first RE, which is used by the second, third, and fifth users.

The uplink SCMA system model is shown in Fig.~\ref{fig:scmasys}. Let $\mathbf{a}_k\in \mathbb{B}^{K_c}$ ($\mathbb{B}=\{0,1\}$ is the set of binary numbers) denote the message word of user $k$ that contains a sequence of independent information bits with the length of $K_c$. 
The message word of each user is fed to its corresponding channel encoder to produce a codeword of length $N_c$.} 
In an $M$-point signal constellation of each user, each $L_M=\log_2 M$ bits are represented by one constellation point.
The codewords {\color{black}from the channel encoders are then divided into $N_{cu} = N_c/L_M$ digital symbols of $L_M$ bits each. In essence, $N_{cu}$ represents the number of channel uses required to transmit one codeword. Let $c_{k,i}\in\mathbb{B}^{L_M}$, $k\in \left\{1,\dots, K\right\}$, $i\in\{1,\dots,N_{cu}\}$ denote the $i$th digital symbol of user $k$.} 
Each symbol of the $k^{\textrm{th}}$ user is then mapped to {\color{black} a $d_v$-dimensional complex constellation point, $\tilde{\mathbf{x}}_{k,i}=\left(\tilde{x}_{1,k,i},\dots,\tilde{x}_{d_v,k,i}\right)^{\textrm{T}}$, which is equivalent to a $2d_v$-dimensional real constellation point,\footnote{From now on, the {\color{black}terms  \lq\lq $d_v$-dimensional complex constellation\rq\rq,~\lq\lq $2d_v$-dimensional real constellation\rq\rq,}~and \lq\lq multidimensional constellation\rq\rq~are used interchangeably.} and is selected from a $d_v$-dimensional complex constellation, $\mathbf{X}_k$, of size $M$. In other words, $\mathbf{X}_k$ is a $d_v\times M$ matrix with each of its column corresponding to $\mathbf{x}_{k,m}=\left(x_{1,k,m},\dots,x_{d_v,k,m}\right)^{\textrm{T}}$, $m\in\{1,\dots,M\}$. For example, in a 4-point signal constellation, the first column of $\mathbf{X}_k$ represent the constellation point corresponds to 00, the second column to 01, the third column to 10, and the fourth column of $\mathbf{X}_k$ corresponds to 11. In that case, $c_{k,i}=00$ is mapped to $\tilde{\mathbf{x}}_{i,k}=\mathbf{x}_{k,1}$, $c_{k,i}=01$ to $\tilde{\mathbf{x}}_{i,k}=\mathbf{x}_{k,2}$, $c_{k,i}=10$ to $\tilde{\mathbf{x}}_{i,k}=\mathbf{x}_{k,3}$, and $c_{k,i}=11$ is mapped to $\tilde{\mathbf{x}}_{i,k}=\mathbf{x}_{k,4}$.

In an SCMA spreading process, each $d_v$-dimensional complex constellation point, $\tilde{\mathbf{x}}_{i,k}$, is \emph{coded} to an sparse $N$-dimensional complex codeword, $\tilde{\mathbf{v}}_{i,k}$, through the binary user-to-RE mapping matrix, $\mathbf{F}_k$. That is, $\tilde{\mathbf{v}}_{i,k}=\mathbf{F}_k\:\tilde{\mathbf{x}}_{i,k}$. The $N$-dimensional complex codeword $\tilde{\mathbf{v}}_{i,k}$ is a sparse vector with $d_v\ll N$ non-zero entries, and is selected from an $N$-dimensional sparse codebook, $\mathbf{V}_k$, of size $M$. In other words, $\mathbf{V}_k$ is an $N\times M$ sparse matrix, and $\mathbf{V}_k=\mathbf{F}_k\:\mathbf{X}_k$.\footnote{{\color{black}To clarify, an \emph{SCMA multidimensional constellation} point is a $d_v$-dimensional complex constellation point without the effect of the user-to-RE mapping matrix. An \emph{SCMA codeword} is an SCMA multidimensional constellation point with the effect of the user-to-RE mapping matrix. Moreover, in an SCMA system with the presence of channel coding (coded SCMA), there are two notions of codewords; one is the binary codeword that is the output of the channel encoder, and the other is the complex codeword which corresponds to the output of the SCMA spreading. The set of the complex codewords is called a codebook.}} Each column of $\mathbf{V}_k$ corresponds to one $N$-dimensional SCMA sparse codeword, $\mathbf{v}_{k,m}=\left(v_{1,k,m},\dots,v_{N,k,m}\right)^{\textrm{T}}$ with $d_v\ll N$ non-zero entries.  The process of SCMA codebook design will be discussed in details in Section \ref{sec:constdesign}.

As shown in Fig.~\ref{fig:scmasys}, the resulant codeword for the $k$th user, $\tilde{\mathbf{v}}_{k,i}$, is then sent over its corresponding channel. The Rayleigh fading channel coefficient vector for the $k$th user over the $i$th channel use is denoted by $\mathbf{h}_{k,i}=\left(h_{1,k,i}, \dots,h_{N,k,i}\right)^{\textrm{T}}$, where $h_{n,k,i}\thicksim\mathcal{CN}\left(0, 1\right)$, $n\in \{1,\dots,N\}$. That is, each channel coefficient, $h_{n,k,i}$, has a complex Gaussian distribution with a mean of 0 and a variance of $E\left[|h_{n,k,i}|^2\right]=1$. Note that since each SCMA user spreads its data over $d_v$ REs out of the $N$ available REs, only $d_v$ of the channel coefficients in $\mathbf{h}_{k,i}$ are relevant. The positions of those $d_v$ channel ceofficeints is according to the positions of the non-zero elements in $\mathbf{S}_k$, and the correlation between them depends on the channel model under study as we will discuss in Section \ref{sec:chnmodel}.

At the receiver in an uplink SCMA system, the $N$-dimensional received column vector corresponding to the $i$th channel use, $\mathbf{y}_i$, can be expressed as
\begin{eqnarray}\label{r}
  \mathbf{y}_i&=&\sum\limits_{k=1}^{K}{\textrm{diag}}\left(\mathbf{h}_{k,i}\right) \tilde{\mathbf{v}}_{k,i}+\bf{w}\nonumber\\
  &=& \sum\limits_{k=1}^{K}{\textrm{diag}}\left(\mathbf{h}_{k,i}\right) \mathbf{F}_k\tilde{\mathbf{x}}_{k,i}+\bf{w},
   \end{eqnarray}}
where $\mathbf{w}\thicksim\mathcal{CN}\left(0,N_0\:\mathbf{I}\right)$ is an $N$-dimensional vector of independent and identically distributed complex Gaussian noise with zero mean and a covariance matrix of $N_0\:\mathbf{I}$. 

Due to the sparsity of SCMA codewords, the {\color{black}widely-used} near-optimal Log-MPA, which is discussed in e.g., \cite{wei2016low}, is used to detect the SCMA codewords. 
 {\color{black}The estimated SCMA codewords are} then used to calculate the bit log-likelihood-ratios (LLRs) {\color{black}(e.g.,\cite{vameghestahbanati2017polar})}, which are fed to each channel decoder to estimate the message word {\color{black}of each user, $\hat{\mathbf{a}}_k$.}
\section{SCMA Multidimensional Constellation Design}\label{sec:constdesign}
\begin{figure}[!t]
\centering
{\includegraphics[width= 0.4\textwidth]{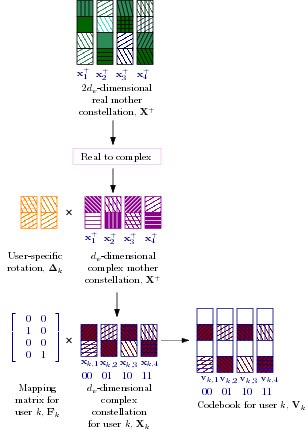}}
\caption{Illustration of the sub-optimal SCMA codebook design approach{\color{black}, with $d_v=4$, $N=4$, $M=4$,} for each user.}
\label{fig:codebook}
\end{figure}

The structure of an SCMA system, $\mathcal{S}$, with $K$ users and $N$ REs, with the user-to-RE mapping matrix of each user, $\mathbf{F}_k$, where each user occupies $d_v$ REs, and employs an $M$-point $N$-dimensional codebook, can be written as \cite{taherzadeh14}
\begin{equation}\label{eqn:struc}
  \mathcal{S}\left(\mathcal{F}, \mathcal{X}; K, M, d_v, N\right),
\end{equation}
where $\mathcal{X} = \left\{\mathbf{X}_k| k=1,\dots,K\right\}$, and $\mathcal{F} = \left[\mathbf{F}_k\right]_{k=1}^{K}$.
The SCMA codebook design problem involves finding the optimum user-to-RE mapping matrix, $\mathcal{F}^\ast$, along with the optimum multidimensional constellation, $\mathcal{X}^\ast$, and can be defined as
\begin{equation}\label{eqn:codebookdesign}
  \mathcal{F}^\ast,\mathcal{X}^\ast=\arg \mathop {\max }\limits_{{\mathcal{F}, \mathcal{X}}}D\left(\mathcal{S}\left(\mathcal{F}, \mathcal{X}; K, M, d_v, N\right)\right),
\end{equation}
where $D$ is a design criterion. A sub-optimal multi-stage optimization approach is deployed in \cite{taherzadeh14,Taherzadeh16patent,zhang16orig,Baoconstellation,cqam17,nikopour2016systems,Bao16sphericalcode,Yan16lattice,peng17,Yan17access,Klimentyevcodebook17,zhai17,Kulhandjian17,He17ldpcscma,Bao18}  for uplink SCMA systems that comprises of finding $\mathcal{F}^{\ast}$ first and then $\mathcal{X}^\ast$. Moreover, when the system is fully loaded, there is a unique solution for $\mathcal{F}^{\ast}=\mathcal{F}$. As such, the optimization problem of an SCMA system is reduced to
\begin{equation}\label{eqn:codebookdesign2}
  \mathcal{X}^\ast=\arg \mathop {\max }\limits_{{\mathcal{X}}}D\left(\mathcal{S}\left({\color{black}\mathcal{F}^{\ast}}, \mathcal{X}; K, M, d_v, N\right)\right).
\end{equation}
The problem is then to find $K$ different $M$-point $d_v$-dimensional complex constellations. To further simplify this optimization problem, the SCMA multidimensional constellation design can be performed in two steps \cite{scma}: Firstly, to design a {\color{black}$d_v\times M$} mother multidimensional constellation $\mathbf{X}^+$ {\color{black}(it follows that each column of $\mathbf{X}^+$ represents one constellation point of the mother constellation, i.e., $\mathbf{x}^{+}_{m}=(x^{+}_{1,m},\dots,x^{+}_{d_v,m})^{\textrm{T}}$, $m\in\{1,\dots,M\}$)}, and secondly, to perform some user-specific unitary rotations on the mother constellation to generate user-specific multidimensional constellations (e.g., \cite{Beek09}). That is, $\mathbf{X}_k={\color{black}\mathbf{\Delta}_k\:\mathbf{X}^+}$, where $\mathbf{\Delta}_k$ represents a {\color{black}$d_v\times d_v$} user-specific rotation matrix. Therefore, $\mathcal{X}=\left\{{\color{black}\mathbf{\Delta}_k\:\mathbf{X}^+}\right\}$. The optimization problem in \eqref{eqn:codebookdesign2} can be expressed as
\begin{equation}
   {\color{black}\left\{\mathbf{\Delta}_k^\ast\right\},\mathbf{X}^{+^\ast}}=\arg \mathop {\max }\limits_{{\color{black}\left\{\mathbf{\Delta}_k\right\},\mathbf{X}^+}}D\left(\mathcal{S}\left({\color{black}\mathcal{F}^{\ast}}, \left\{{\color{black}\mathbf{\Delta}_k\:\mathbf{X}^+}\right\}; K, M, d_v, N\right)\right).
\end{equation}
As a suboptimal approach, the mother constellation and the user-specific rotations {\color{black}can be} found separately \cite{taherzadeh14}. {\color{black}More particularly, an $M$-point $2d_v$-dimensional real mother constellation, $\mathbf{X}^+$, is first designed (according to the criteria that will be discussed in details in Section \ref{sec:chnmodelKPI}), and then converted to an $M$-point $2d_v$-dimensional complex constellation {\color{black}using a real-to-complex converter}. The $d_v\times d_v$ specific rotation for user $k$, $\mathbf{\Delta_k}$, is applied on the complex $\mathbf{X}^+$ to create $\mathbf{X}_k$. The allocation of REs to the $k^{\textrm{th}}$ user is performed through $\mathbf{F}_k$, which is then applied on the $d_v$-dimensional complex constellation, $\mathbf{X}_k$, to create the codebook $\mathbf{V}_k$.

 Fig.~\ref{fig:codebook} illustrates the sub-optimal process of designing a 4-point SCMA codebook for user $k$ with $N=4$, $d_v=2$, and $\mathbf{F}_k$ in \eqref{fk}. The 4-point 4-dimensional real mother constellation points, $\{\mathbf{x}^{+}_{m}\}$, $m\in\{1,\dots,M\}$, are labeled by $\mathbf{x}^{+}_{1}$ to $\mathbf{x}^{+}_{4}$. Each dimension of the 4-dimensional $\mathbf{x}^{+}_{m}$ is a representation of $x^{+}_{j',m}$, $j'\in\{1,\dots,2d_v\}$, and is shown by a square with a specific pattern. The 4-dimensional real mother constellation $\mathbf{X}^+$ is first designed and then converted to a 4-point 2-dimensional complex mother constellation. In Fig.~\ref{fig:codebook}, each square of the 2-dimensional complex $\mathbf{X}^+$ is a representation of $x^+_{j,m}$, $j\in\{1,\dots,d_v\}$, $d_v=2$. The $2\times 2$ specific rotation for user $k$, i.e., $\mathbf{\Delta_k}$, is then applied on the complex $\mathbf{X}^+$ to create the 4-point 2-dimensional complex constellation for user $k$, i.e., $\mathbf{X}_k$. The points of $\mathbf{X}_k$ are labeled by $\mathbf{x}_{k,1}$, $\mathbf{x}_{k,2}$, $\mathbf{x}_{k,3}$, $\mathbf{x}_{k,4}$ and tagged by their corresponding digital symbols, 00, 01, 10, and 11. That is, 00 is mapped to $\mathbf{x}_{k,1}$, 01 to $\mathbf{x}_{k,2}$, 10 to $\mathbf{x}_{k,3}$, and 11 is mapped to $\mathbf{x}_{k,4}$. Each dimension of the 2-dimensional $\mathbf{x}_{k,m}$, i.e., $x_{j,m}$, in Fig.~\ref{fig:codebook} is represented by a square with a certain pattern. The binary mapping matrix, $\mathbf{F}_k$ is applied on the 2-dimensional complex constellation, $\mathbf{X}_k$, to create the 4-dimensional complex codebook, $\mathbf{V}_k$. Each complex codeword in Fig.~\ref{fig:codebook} is labeled by $\mathbf{v}_1$ to $\mathbf{v}_4$, tagged by 00, 01, 10, 11, and represented by four squares. The four squares correspond to the four REs ($N=4$). That is, each square represents $v_{n,k,m}$, $n\in\{1,\dots,4\}$, $k\in\{1,\dots,6\}$, $m\in\{1,\dots,4\}$.}  

Note that the primary purpose of user-specific rotations is to maintain uniquely decodable symbols for users that collide at the same RE. {\color{black}This is very important in downlink scenarios, but} in uplink scenarios the channel {\color{black}also} introduces random rotations to constellations, {\color{black}in which case} the optimization of 2-dimensional user-specific rotations becomes questionable\footnote{{\color{black}More details on the effect of user-specific rotations are provided in Appendix \ref{app:userspecific}.}}. That being said, the optimization of the mother constellation $\mathbf{X}^+$ has a significant impact on the performance of the SCMA system.

In the following section, we highlight the key performance indicators (KPIs) of $M$-point multidimensional constellations {\color{black}for SCMA systems that should be considered in their design process} under different channel scenarios. We then provide an overview of different approaches that have been performed to design SCMA multidimensional constellations in Section \ref{sec:constellations}.

\section{Channel Models and Key performance indicators}\label{sec:chnmodelKPI}
As SCMA has been proposed in a variety of contexts, we provide different channel scenarios under study in Section \ref{sec:chnmodel}. We
then review the KPIs that should be considered in the design process of $M$-point multidimensional mother constellations in Section \ref{sec:kpi}.

\subsection{Channel Models}\label{sec:chnmodel}
{\color{black}In uplink SCMA systems, where each user spreads its data over $d_v$ REs, the channel can be categorized based on i) the correlation between the fading coefficients of the different REs at each user, and ii) the temporal correlation of the fading coefficients at each RE at different times. In the first category, at one extreme we have fully correlated  coefficients (that is, all the coefficients for a user are the same, so no diversity is attained), which typically occurs when the REs are adjacent, and at the other extreme, we have fully independent coefficient, which typically occurs when the REs are far apart. In the second category, which is particularly relevant to systems that use channel coding, at one extreme we have \lq\lq fast\rq\rq~fading, where fully independent channel coefficients are observed with each channel use, and at the other extreme we have slow \lq\lq quasi-static\rq\rq~fading, where the channel coefficients remain constant for the duration of the transmission of an entire codeword.}

 {\color{black}Recall from Section \ref{sec:sysmodel} that the channel coefficient for the $k$th user over the $n$th RE and for the $i$th channel use is denoted by $h_{n,k,i}$}. The channel models under study are as follows:
\begin{list}{Case~\arabic{qcounter}:}{\usecounter{qcounter}}
  \item Uncoded fading where each user observes the same channel coefficients over their REs (FSC). {\color{black}More precisely, since in uncoded systems there is only one use of the channel, i.e., $N_{cu}=1$, $E[h_{n,k,1}\:{h^{*}}_{n',k,1}]=1$, for $n,\:n'\in\{1,\dots,N\}$.} \label{case:fsc}
  \item Uncoded fading where each user observes {\color{black} independent} channel coefficients over their REs ({\color{black}FIC}). {\color{black}More precisely, $E[h_{n,k,1}\:{h^{*}}_{n',k,1}]=\delta[n-n']$, where $\delta[\cdot]$ denotes the Kronecker delta function.} \label{case:fdc}
  \item Coded fast fading where each user observes the same channel coefficients over their REs (FFSC). {\color{black}More precisely, $E[h_{n,k,i}\:{h^{*}}_{n',k,i'}]=\delta[i-i']$, $i,\:i'\in\{1,\dots,N_{cu}\}$.}\label{case:ffsc}
  \item Coded fast fading where each user observes {\color{black} independent} channel coefficients over their REs ({\color{black}FFIC}). {\color{black}More precisely, $E[h_{n',k,i}\:{h^{*}}_{n',k,i'}]=\delta[n-n']\:\delta[i-i']$.}  \label{case:ffdc}
  \item Coded {\color{black} quasi-static} fading where each user observes the same channel coefficients over their REs (SFSC). {\color{black}More precisely, $E[h_{n,k,i}\:{h^{*}}_{n',k,i'}]=1$.}  \label{case:sfsc}
  \item Coded {\color{black} quasi-static} fading where each user observes {\color{black} independent} channel coefficients over their REs ({\color{black}SFIC}). {\color{black}More precisely, $E[h_{n,k,i}\:{h^{*}}_{n',k,i'}]=\delta[n-n']$.} \label{case:sfdc}
\end{list}

\subsection{KPIs}\label{sec:kpi}
In this section, we highlight the KPIs that should be considered in the design process of an $M$-point $2d_v$-dimensional mother constellation, $\mathbf{X}^+$, under different channel scenarios.
For notational brevity, we drop the superscript $+$ from $\mathbf{X}^+$, and denote a $2d_v$-dimensional real mother constellation by $\mathbf{X}$. Each constellation point is then represented by $\mathbf{x}_m$.

\begin{figure}[!t]
\centering
\includegraphics[trim = 5mm 20mm 10mm 20mm, clip,width=.4\textwidth]{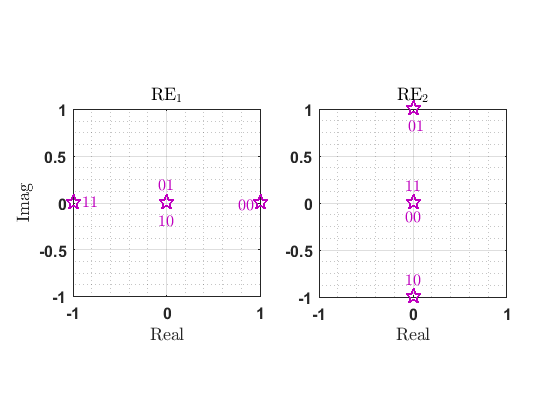}
\caption{{\color{black}The projection of a 4-point 4-dimensional constellation onto the first pre-assigned RE ($\textrm{RE}_1$), and the second pre-assigned RE ($\textrm{RE}_2$).}}
\label{fig:exp}
\end{figure}

Multidimensional constellations can be projected onto a set of orthogonal 2-dimensional signal spaces (i.e., the I and Q channels) where each projection can be transmitted independently (e.g., \cite{khoshnevis17}). We represent the projection of the $j$th complex element of $\mathbf{x}_m$ onto two dimensions by $\mathbf{x}_{mj}$, $1\leq j\leq d_v$. For a scenario with $d_v=2$, consider the 4-point 4-dimensional constellation in Table \ref{table:4CQAM}. As we see, $00$ is mapped to $(1,0,0,0)$, $01$ to $(0,0,0,1)$, $10$ to $(0,0,0,-1)$, and $11$ to $(-1,0,0,0)$. The projection of each constellation point onto the first pre-assigned RE ($\textrm{RE}_1$) and onto the second pre-assigned RE ($\textrm{RE}_2$) are denoted by $x_{m1}$ and $x_{m2}$ in Table \ref{table:4CQAM}, and depicted in Fig.~\ref{fig:exp}. 

The average symbol energy of a constellation is defined as
\begin{equation}\label{eqn:Es}
  E_s= \frac{1}{M}\sum_{m=1}^{M}\lVert \mathbf{x}_m \rVert^2.
\end{equation}
In this paper, we have normalized the energies of all constellations to unity in order to provide a consistent comparison.
\begin{table}[!t]
\caption{An example of a 4-point 4-dimensional constellation} 
\centering 
\begin{tabular}{cccc } 
\multicolumn{4}{c}{} \\ [0.7ex]
 \hline 
  &$\mathbf{x}_m$&${x}_{m1}$&${x}_{m2}$\\ 
 \hline
 $00$ &$(1,0,0,0)$ &$1+0i$& $0+0i$ \\
 $01$ & $(0,0,0,1)$ &$0+0i$& $0+1i$ \\
  $10$ & $(0,0,0,-1)$ &$0+0i$& $0-1i$ \\
  $11$ & $(-1,0,0,0)$ &$-1+0i$& $0+0i$ \\
 \hline
\end{tabular}
\label{table:4CQAM} 
\end{table}
 The KPIs of $M$-point $2d_v$-dimensional constellations that significantly impact the performance of SCMA systems over different channel scenarios are as follows:
\subsubsection{Euclidian Distance}
The Euclidian distance between two constellation points, $\mathbf{x}_m$ and $\mathbf{x}_{m'}$, $1\leq m < m'\leq M$, is
\begin{equation}\label{dE}
  {\color{black}d_{E}^{mm'}} = \left\lVert\mathbf{x}_m-\mathbf{x}_{m'}\right\rVert,
\end{equation}
The minimum Euclidian distance of a constellation is defined as
\begin{equation}\label{dpmin}
  d_{E,\:\min} = \min \left\{{d_{E}^{mm'}}\mid 1\leq m < m'\leq M\right\}.
\end{equation}
It is known that the minimum Euclidian distance of a constellation is a KPI in AWGN channels \cite{Proakis08,Beko12,Biglieri98}. Moreover, in fading scenarios with $d_v$ adjacent REs (dependent fading) 
since the constellation points over the $d_v$ REs are rotated and scaled by the same values (the same channel coefficients), the relative Euclidian distance between the constellation points remains intact. {\color{black}As such, $d_{E,\:\min}$ is an important factor in FSC, FFSC, SFSC.}
\subsubsection{Euclidian Kissing Number}
The Euclidian kissing number, $\tau_E$, is defined as the average number of constellation pairs at the minimum Euclidian distance $d_{E,\:\min}$, and is a KPI in FSC, FFSC, and SFSC. 
\subsubsection{Product Distance}\label{pdist}
The product distance between two $d_v$-dimensional complex constellation points, $\mathbf{x}_m$ and $\mathbf{x}_{m'}$, is defined as \cite{Proakis08}
\begin{equation}\label{eqn:dpmin}
  {\color{black}{d_{P}^{mm'}}} = \prod_{j\in {\mathcal{J}}_{mm'}} \left| x_{mj}-x_{m'j} \right|,
\end{equation}
where $x_{mj}$ and $x_{m'j}$ are the $j$th {\color{black}complex} element of $\mathbf{x}_m$ and $\mathbf{x}_{m'}$, respectively. Moreover, ${\mathcal{J}}_{mm'}$ denotes the set of dimensions, $j$, for which $x_{mj}\neq x_{m'j}$, i.e.,
\begin{equation}\label{jmm}
  {\mathcal{J}}_{mm'} = \left\{j\mid x_{mj}\neq x_{m'j}, 1\leq j \leq d_v \right\}.
\end{equation}
The minimum product distance is
\begin{equation}\label{dpmin}
  d_{P,\:\min} = \min \left\{{d_{P}^{mm'}}\mid 1\leq m < m'\leq M\right\}.
\end{equation}
It is known from \cite{Boutros96,Boutros98,Biglieri98} that maximizing the minimum product distance of the constellation points has an important impact on the performance of the system in fading scenarios with $d_v$ non-adjacent REs (independent fading), i.e., FIC, FFIC, SFIC, especially at high SNRs.
\subsubsection{Product Kissing Number}
The product kissing number, $\tau_P$, is defined as the average number of constellation pairs at the minimum product distance $d_{P,\:\min}$. It is shown in \cite{Boutros96} that $\tau_P$ is one of the dominant factors in the symbol error probability of a multidimensional constellation in fading scenarios with $d_v$ non-adjacent REs, i.e., FIC, FFIC, SFIC.
\subsubsection{Modulation Diversity Order}
The modulation/signal space diversity order, $L$, of a multidimensional constellation is defined as the minimum number of distinct components between any two constellation points \cite{Boutros98}. In other words, $L$ is the minimum Hamming distance between any two different constellation points. That is,
\begin{equation}\label{L}
  L = \min \left\{{d_H\left(\mathbf{x}_m,\mathbf{x}_{m'}\right)}\mid 1\leq m < m'\leq M\right\},
\end{equation}
{\color{black}where ${d_H\left(\mathbf{x}_m,\:\mathbf{x}_{m'}\right)}$ represents the \emph{Hamming distance} between $\mathbf{x}_m$ and $\mathbf{x}_{m'}$, i.e., the cardinality of ${\mathcal{J}}_{mm'}$ in \eqref{jmm}.}
It is known (e.g., \cite{Boutros98,Biglieri98,Chindapol01}) that $L$ plays an important role in fading channels with $d_v$ non-adjacent REs, i.e., FIC, FFIC, SFIC.\footnote{{\color{black}It is worthwhile to note that the fading channel diversity order is equal to $L$ in fading channels with $d_v$ non-adjacent REs, i.e., FIC, FFIC, SFIC, whereas the fading channel diversity order is equal to 1 in fading channels with $d_v$ adjacent REs, i.e., FSC, FFSC, SFSC.}}
\subsubsection{Number of Distinct Points}
{\color{black}In SCMA systems, the average number of distinct points, $N_d$, $0<N_d\leq M$, along the projections of an $M$-point $2d_v$-dimensional constellation onto two dimensions (I and Q channels) is a KPI in most scenarios, 
and is defined as
{\color{black}\begin{equation}\label{Nd}
  N_d = \frac{1}{d_v}\sum_{j=1}^{d_v}\sum_{m=1}^{M}\delta\left[\sum_{m'< m}C\left(\left|{x}_{mj}-{x}_{m'j}\right|\right)\right],
\end{equation}}
where
\begin{eqnarray}
C\left( z \right) = \left\{ {\begin{array}{*{20}c}
   1 & {\textrm{if}~z = 0}  \\
   0 & {\textrm{if}~z \ne 0}  \\
\end{array}}, \right.
\end{eqnarray}
{\color{black}and {\color{black}$\delta[\cdot]$} denotes the Kronecker delta function.}

 For the constellation provided in  Table \ref{table:4CQAM} and depicted in Fig.~\ref{fig:exp}, it is clear that both $01$ and $10$ overlap over $\textrm{RE}_1$, and both $00$ and $11$ overlap over $\textrm{RE}_2$. Referring to \eqref{Nd}, for this constellation $N_d=3$.

  As we will discuss later in Section \ref{sec:sim}, in general, the higher $N_d$, the better the performance. That is, if there is an overlap among different symbols, i.e., $N_d\neq M$, in a scenario where one of the REs is in a deep fade, the transmitted symbol cannot easily be recovered from other REs.}
\subsubsection{Bit-labeling}
In uncoded systems, the mapping function that maps the incoming bits to each constellation point, i.e., bit-labeling, has an impact on the BER performance. However, bit-labeling has no impact on the symbol error performance (SER) of uncoded systems \cite{Proakis08}. In {\color{black} coded systems}, bit-labeling plays an important role on the system performance in different channel scenarios (e.g., \cite{Reinhard17}). This can be due to the fact that in {\color{black} many} coded systems, the output of the detector is used to calculate the \emph{bit} LLRs, which are then fed to the channel decoder. As such, the BER performance of the uncoded system gives some indication of the system performance in the presence of the channel coding.

Bit-labeling for BICM has been studied widely e.g., \cite{Rachinger17, Stierstorfer07,Agrell09,Agrell11,Chindapol01,Alamri10,Caire98}. In \cite{Caire98}, it is conjectured that Gray labeling maximizes the capacity of BICM. In Gray labeling, two points in the constellations {\color{black}that} are adjacent (in terms of the Euclidian distance) {\color{black}must} differ in {\color{black}only} one bit position.
The multidimensional constellations based on hypercubes, e.g., constellations constructed similar to \cite{Boutros98}, with Gray labeling provide an equal error protection among their different bit levels. As such, they perform well when employed in conjunction with BICM. Note that by Gray labeling of the $2d_v$-dimensional constellation, we mean Gray labeling in the $2d_v$-dimensional space, and not necessarily Gray labeling along the projection of the constellation along two dimensions. 

In addition to the KPIs mentioned above, the behavior of the multi-user detector at different SNR regions can accordingly effect the behavior of constellations in different scenarios. Moreover, the code rate has a significant impact on the system performance. 
\section{Overview of SCMA Multidimensional Constellations}\label{sec:constellations}
 In this section, we provide an overview of the $M$-point $2d_v$-dimensional real constellations for uplink SCMA systems proposed in 
 \cite{hoshyar08,taherzadeh14,Taherzadeh16patent,zhang16orig,Baoconstellation,cqam17,nikopour2016systems,Bao16sphericalcode,Yan16lattice,peng17,Yan17access,Klimentyevcodebook17
,zhai17,Kulhandjian17,He17ldpcscma,Bao18} for $M\in\{4,16\}$, and $d_v=2$. That is, each user sends its data over two REs only\footnote{Note that most of the constellations proposed for SCMA systems are given for $M\in\{4,16\}$, $d_v=2$. As an essential purpose of SCMA systems is to support massive connectivity in IoT applications \cite{Bai17IoT}, and not increasing the throughput of individual users, $M\in\{4,16\}$ is a fair assumption. In addition, as the user-to-RE indicator matrix for an SCMA system {\color{black}is required} to be a sparse matrix, $d_v=2$ is an important scenario.}.
\begin{figure*}
\centering     
\subfigure[]{\label{fig:4lds}\includegraphics[trim = 5mm 20mm 10mm 20mm, clip,width=.4\textwidth]{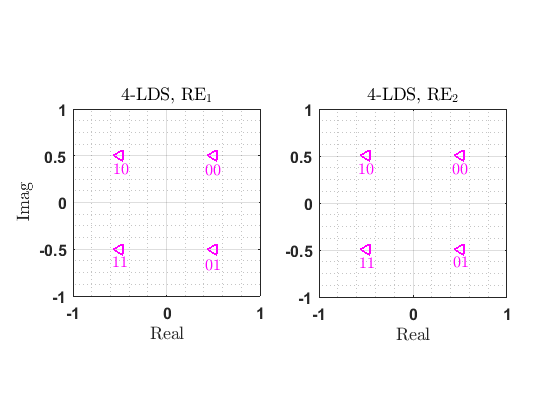}}
\subfigure[]{\label{fig:T4qam}\includegraphics[trim = 5mm 20mm 10mm 20mm, clip,width=.4\textwidth]{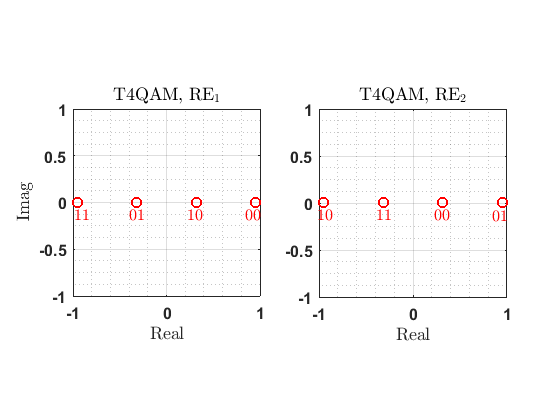}}
\subfigure[]{\label{fig:4scma}\includegraphics[trim = 5mm 20mm 10mm 20mm, clip,width=.4\textwidth]{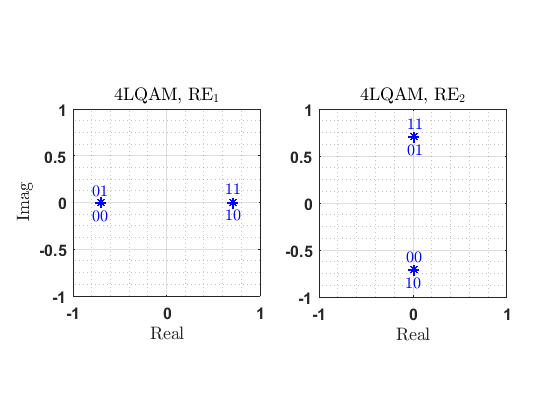}}
\subfigure[]{\label{fig:4bao}\includegraphics[trim = 5mm 20mm 10mm 20mm, clip,width=.4\textwidth]{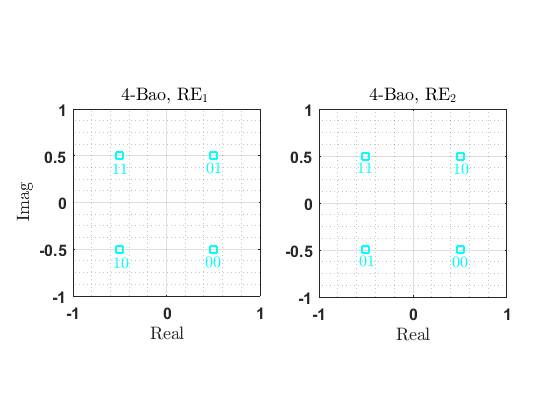}}
\subfigure[]{\label{fig:4cqam}\includegraphics[trim = 5mm 20mm 10mm 20mm, clip,width=.4\textwidth]{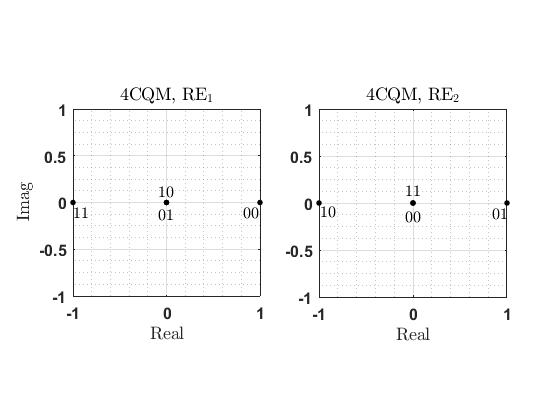}}
\subfigure[]{\label{fig:4beko}\includegraphics[trim = 5mm 20mm 10mm 20mm, clip,width=.4\textwidth]{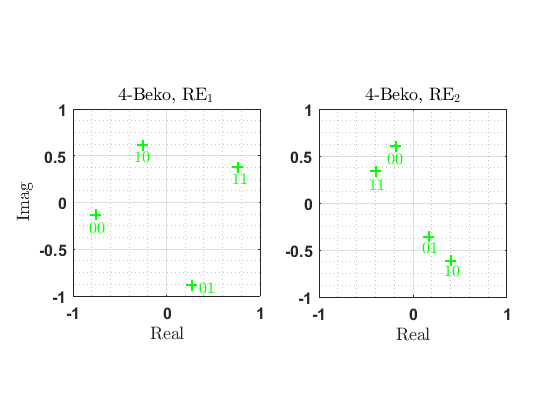}}
\caption{The projection of 4-point 4-dimensional constellations over $\textrm{RE}_1$ and $\textrm{RE}_2$: (a) 4-LDS, (b) T4QAM, (c) 4LQAM, (d) 4-Bao, (e) 4CQAM, and (f) 4-Beko.}
\label{fig:4aryconst}
\end{figure*}
\begin{figure*}
\centering     
\subfigure[]{\label{fig:16ldsRE1}\includegraphics[width= 0.45\textwidth]{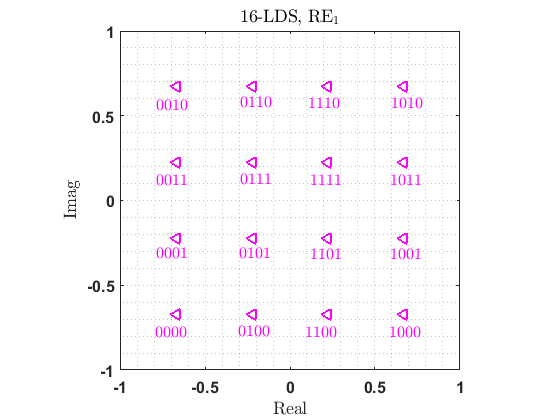}}
\subfigure[]{\label{fig:16ldsRE2}\includegraphics[width= 0.45\textwidth]{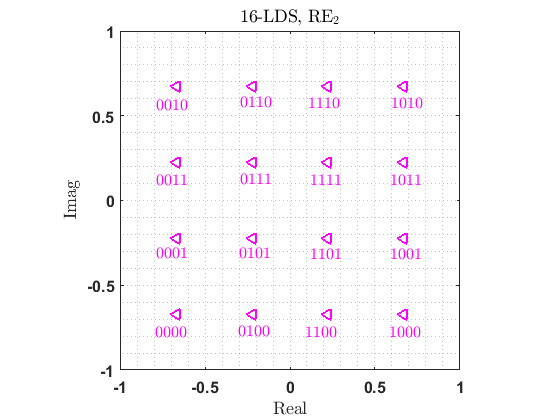}}
\subfigure[]{\label{fig:T16qamRE1}\includegraphics[width= 0.45\textwidth]{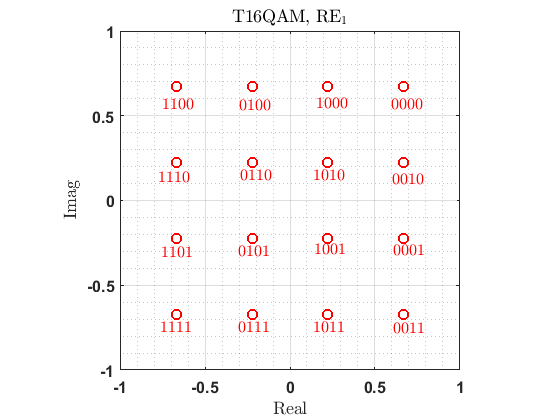}}
\subfigure[]{\label{fig:T16qamRE2}\includegraphics[width= 0.45\textwidth]{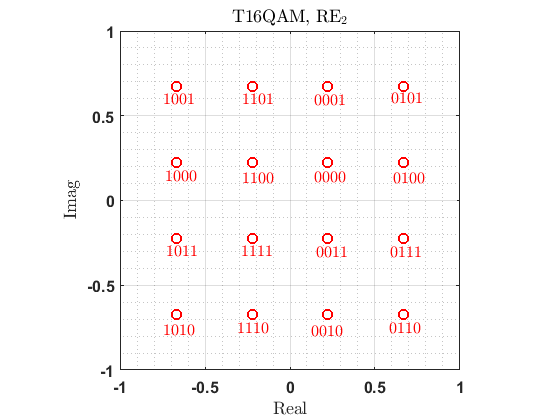}}
\subfigure[]{\label{fig:16LqamRE1}\includegraphics[width= 0.45\textwidth]{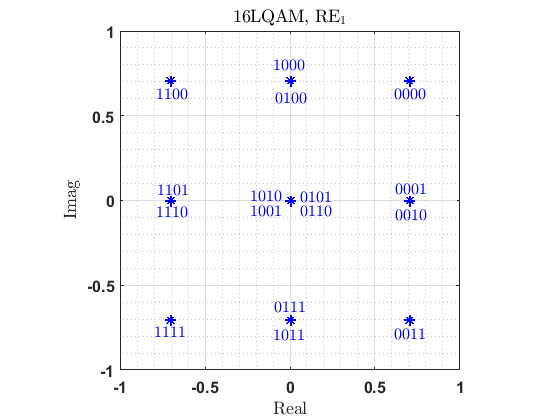}}
\subfigure[]{\label{fig:16LqamRE2}\includegraphics[width= 0.45\textwidth]{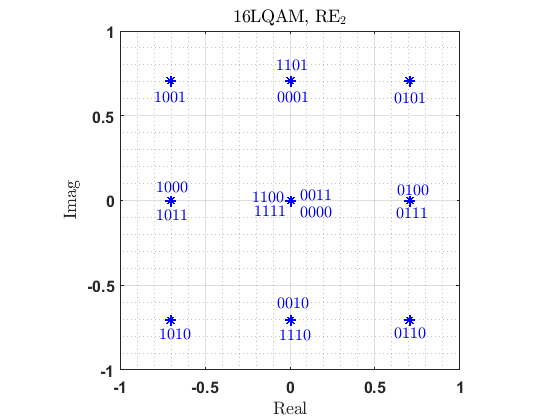}}
\caption{The projection of 16-point 4-dimensional constellations over $\textrm{RE}_1$ and $\textrm{RE}_2$: (a) 16-LDS over $\textrm{RE}_1$, (b) 16-LDS over $\textrm{RE}_2$, (c) T16QAM over $\textrm{RE}_1$, (d) T16QAM over $\textrm{RE}_2$, (f) 16LQAM over $\textrm{RE}_1$, and (f) 16LQAM over $\textrm{RE}_2$.}
\label{fig:16const1}
\end{figure*}

The $M$-point $2d_v$-dimensional constellations under study are as follows:
\subsubsection{$M$-LDS}\label{const:MLDS}
The repetition of {\color{black}an} $M$-QAM constellation over the 2 REs, which is referred as $M$-LDS \cite{hoshyar08}. In Fig.~\ref{fig:4lds}, we show the projection of 4-LDS with magenta left-pointing triangles over $\textrm{RE}_1$ and $\textrm{RE}_2$. As an example, with a Gray labeling, to transmit 11 a user would send $(-1/2, -1/2)$ over both $\textrm{RE}_1$ and $\textrm{RE}_2$. Likewise, we depict the projection of 16-LDS with magenta left-pointing triangles over $\textrm{RE}_1$ and $\textrm{RE}_2$ in Fig.~\ref{fig:16ldsRE1} and Fig.~\ref{fig:16ldsRE2}, respectively.
\subsubsection{T$M$QAM}\label{const:TMqam}
The constellation proposed in \cite{taherzadeh14, Taherzadeh16patent}. Inspired by \cite{Boutros98}, the design process of this constellation involves a shuffling method that establishes a $d_v$-dimensional complex constellations from the Cartesian product of two $d_v$-dimensional real constellations with a desired Euclidian distance profile. A unitary rotation matrix is then applied to maximize the minimum product distance,  $d_{P,\:\min}$, of the $2d_v$-dimensional constellation. {\color{black}This $M$-point constellation was named as T$M$QAM when it was proposed in  \cite{taherzadeh14, Taherzadeh16patent}.}

We show the projections of T4QAM onto two dimensions with red circles in Fig.~\ref{fig:T4qam}. Each user maps the incoming bitstreams onto the constellation with red circles in Fig.~\ref{fig:T4qam} in order to send them over $\textrm{RE}_1$ and $\textrm{RE}_2$. For example, to transmit 11, the user would send $(-3/\sqrt{10},0)$ on its first RE and $(-1/\sqrt{10},0)$ on its second RE. Furthermore, we show the projections of T16QAM onto two dimensions for the two REs with red circles in Fig.~\ref{fig:T16qamRE1}--Fig.~\ref{fig:T16qamRE2}. As an example, with a Gray labeling to transmit 1111, the user sends $(-3\sqrt{5}/10,-3\sqrt{5}/10)$ and $(-\sqrt{5}/10,-\sqrt{5}/10)$ on its $\textrm{RE}_1$ and its $\textrm{RE}_2$, respectively.  

It is worth to note that the mother constellation of the widely used (e.g., \cite{Dai16_polar,Sergienko16,Klimentyev16detection,vameghestahbanati2017polar,Dai17TVT,meSD1CL17}) SCMA codebook that is  represented in \cite{codebook}, and later described in \cite{zhang16orig}, is in fact the T4QAM constellation. However, in \cite{zhang16orig,codebook} the 2-dimensional components of the mother constellation are then rotated according to the Latin square criterion \cite{latin91} to obtain multidimensional constellations that are different for different users. Furthermore, the 4-point mother constellation that is proposed in \cite{Yan16lattice} is similar to T4QAM, and is rotated according to \cite{Beek09} to differentiate the users.
As mentioned earlier, the value of the 2-dimensional user-specific rotations becomes questionable in uplink scenarios. In addition, the codebook that is proposed in \cite{Yan17access} is based on the technique devised in \cite{Boutros98}, {\color{black}and} is expected to result in a constellation similar to T$M$QAM. 
      \begin{figure*}
\centering     
\subfigure[]{\label{fig:16baoRE1}\includegraphics[width= 0.45\textwidth]{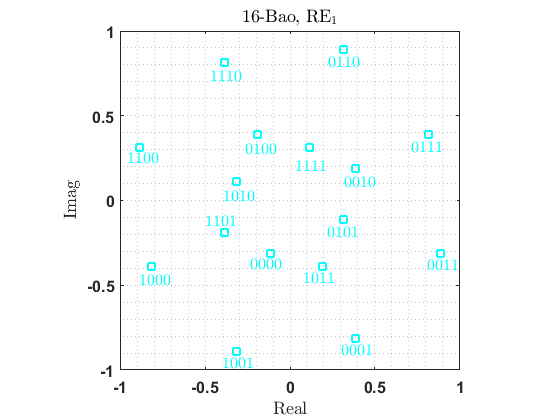}}
\subfigure[]{\label{fig:16baoRE2}\includegraphics[width= 0.45\textwidth]{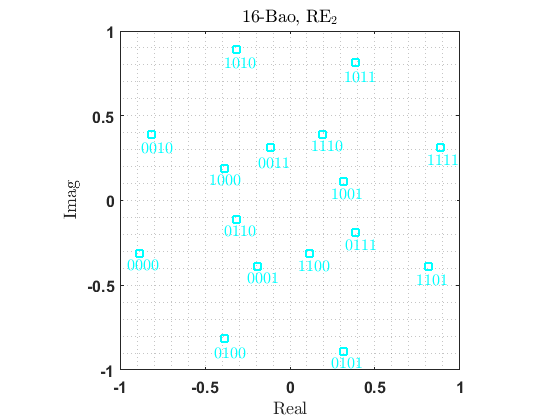}}
\subfigure[]{\label{fig:16HqamRE1}\includegraphics[width= 0.45\textwidth]{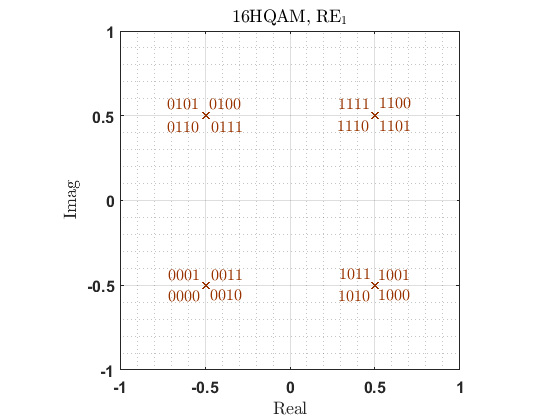}}
\subfigure[]{\label{fig:16HqamRE2}\includegraphics[width= 0.45\textwidth]{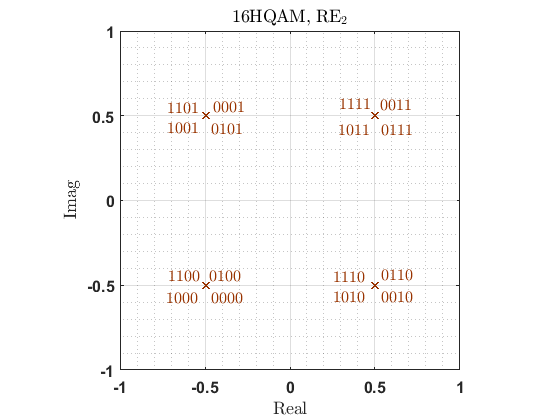}}
\subfigure[]{\label{fig:16CqamRE1}\includegraphics[width= 0.45\textwidth]{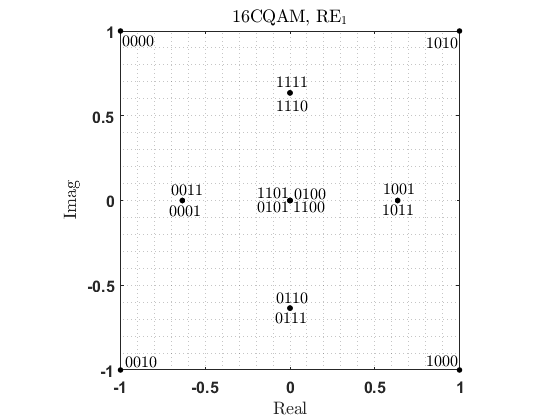}}
\subfigure[]{\label{fig:16CqamRE2}\includegraphics[width= 0.45\textwidth]{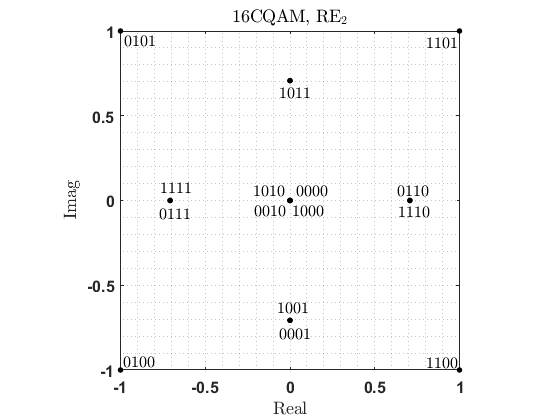}}
\caption{The projection of 16-point 4-dimensional constellations over $\textrm{RE}_1$ and $\textrm{RE}_2$: (a) 16-Bao over $\textrm{RE}_1$, (b) 16-Bao over $\textrm{RE}_2$, (c) 16HQAM over $\textrm{RE}_1$, (d) 16HQAM over $\textrm{RE}_2$, (e) 16CQAM over $\textrm{RE}_1$, and (f) 16CQAM over $\textrm{RE}_2$.}
\label{fig:16const2}
\end{figure*}
\subsubsection{$M$LQAM}\label{const:LMqam}
The constellation with a low number of projection that is proposed in \cite{taherzadeh14, Taherzadeh16patent}, and given in \cite{cqam17}. We refer to the $M$-point constellation in \cite{taherzadeh14, Taherzadeh16patent} as $M$LQAM (QAM with {\color{black}a} \emph{low} number of projections). Similar to T$M$QAM, the design process of $M$LQAM associates with a shuffling method that constitutes a $d_v$-dimensional complex constellations from the Cartesian product of two $d_v$-dimensional real constellations. That said, unlike T$M$QAM, a unitary rotation matrix is then applied to \emph{lower} the number of projected points over each RE, which, in turn, decreases the computational complexity of the detector \cite{bayesteh15}.

We show the projections of 4LQAM onto two dimensions over $\textrm{RE}_1$ and $\textrm{RE}_2$ with blue asterisks in Fig. \ref{fig:4scma}. As we see in Fig.~\ref{fig:4scma}, 4 points of the constellation are mapped to $N_d=2$ points only. That is, with Gray labeling, both 00 and 01 are mapped to $(-\sqrt{2}/2,0)$, and both 10 and 11 are mapped to $(\sqrt{2}/2,0)$ over $\textrm{RE}_1$. Moreover, both 00 and 10 are mapped to $(0,-\sqrt{2}/2)$, and both 01 and 11 are mapped to $(0,\sqrt{2}/2)$ over $\textrm{RE}_2$. As such, the MPA detection computational complexity \cite{taherzadeh14} can be reduced from $4^{d_c}$ to $2^{d_c}$. {\color{black}The 4LQAM constellation in Fig. 4(c) can be considered as the transmission of two independent BPSK constellations on each RE.}
In a similar vein, we depict the projections of 16LQAM onto two dimensions over $\textrm{RE}_1$ and $\textrm{RE}_2$ with blue asterisks in Fig. \ref{fig:16LqamRE1} and Fig. \ref{fig:16LqamRE2}, respectively. It is clear that 16 points of the constellation are mapped to only $N_d=9$ points in each 2-dimensional component. Therefore, the MPA detection computational complexity  can be reduced from $16^{d_c}$ to $9^{d_c}$. Nevertheless, as we will discuss in Section \ref{sec:sim}, 4LQAM and 16LQAM introduce performance degradations in some scenarios. 

  \subsubsection{$M$-Bao}\label{const:Mbao}
The constellation proposed for SCMA systems over Rayleigh fading channels with $d_v$ non-adjacent REs, i.e., FIC (Case \ref{case:fdc}), in \cite{Baoconstellation} that is based on the rotation of QAM constellations. The rotation matrices are obtained through computer search inspired by the approaches in \cite{xin2003spacetime, sanjeewa12} in order to maximize the cutoff rate of the equivalent MIMO systems. This exhaustive search is only feasible if $d_c\leq 3$. We name the $M$-point constellation proposed in \cite{Baoconstellation} as $M$-Bao.

We depict the projection of four-dimensional 4-Bao onto two dimensions over $\textrm{RE}_1$ and $\textrm{RE}_2$ with cyan squares in Fig. \ref{fig:4bao}. With a Gray labeling, to send 01, the user would send $(0.5019, 0.4981)$ over $\textrm{RE}_1$, and $(-0.5019, -0.4981)$ over $\textrm{RE}_2$. Further, we show the projection of four-dimensional 16-Bao onto two dimensions over $\textrm{RE}_1$ and $\textrm{RE}_2$ with cyan squares in Fig. \ref{fig:16baoRE1} and Fig. \ref{fig:16baoRE2}, respectively. For instance, with a Gray labeling, to send 0001, the user would send $(0.3876, - 0.8148)$ and $(-0.1890, - 0.3876)$ over $\textrm{RE}_1$ and $\textrm{RE}_2$, respectively.

In \cite{Bao16sphericalcode}, the spherical codes are proposed for SCMA systems that are based on the constellations in \cite{wong10,sloane81}. It was later shown in \cite{Baoconstellation} by similar authors that $M$-Bao in \cite{Baoconstellation} outperforms the spherical codes in \cite{Bao16sphericalcode}.
\subsubsection{16HQAM}\label{const:16Hqam}
By investigating T16QAM, 16LQAM, and 16-Bao, we notice they all are based on the Cartesian product of two 4-QAM constellations that are rotated using different unitary rotations to fulfill different aforementioned requirements. The Cartesian product of two 4-QAM constellations constitutes the 16 corners of the four-dimensional hypercube. We name the 16-point four-dimensional {\color{black}\emph{hypercube}-based QAM} as 16HQAM, and depict its projection onto two dimensions over $\textrm{RE}_1$ and $\textrm{RE}_2$ with brown crosses in Fig. \ref{fig:16HqamRE1} and Fig. \ref{fig:16HqamRE2}, respectively. As we see, the 16 points of the constellations are mapped to $N_d=4$ points only, which can result in a reduction of MPA detection complexity from $16^{d_c}$ to $4^{d_c}$. That said, as we discuss in Section \ref{sec:sim}, the performance of H16QAM will be penalized in some scenarios by its low number of distinct points. 
\subsubsection{$M$CQAM}\label{const:MCqam}
The circular QAM constellation proposed in \cite{cqam17}. We refer to $M$-point \emph{circular} QAM constellation as $M$CQAM. $M$CQAM is based on the analysis of the signal-space diversity for MIMO systems over Rayleigh fading channels in \cite{Cai16,Baoconstellation}, but it also obtains a low number of projections per complex dimension.

We show the projection of 4CQAM\footnote{This constellation is also provided in \cite{huawei}.} over $\textrm{RE}_1$ and $\textrm{RE}_2$ with black points in Fig.~\ref{fig:4cqam}. As we see, the 4 points of the constellation are mapped to $N_d=3$ points. More specifically, with a Gray labeling, in order to transmit over $\textrm{RE}_1$, both 01 and 10 are mapped to $(0,0)$, 00 is mapped to $(1,0)$, and 11 is mapped to $(-1,0)$. On the other hand, to transmit over $\textrm{RE}_2$, both 00 and 11 are mapped to $(0,0)$, 01 is mapped to $(1,0)$, and 10 is mapped to $(-1,0)$. Hence, the MPA detection computational complexity can be reduced from $4^{d_c}$ to $3^{d_c}$. 
Similarly, we depict the projection of 16CQAM over $\textrm{RE}_1$ and $\textrm{RE}_2$ with black points in Fig.~\ref{fig:16CqamRE1} and Fig.~\ref{fig:16CqamRE2}, respectively. We observe that the 16 points of the constellation are mapped to $N_d=9$ points only. The MPA detection computation is proportional to $9^{d_c}$ instead of $16^{d_c}$. However, akin to $M$LQAM and 16HQAM, $M$CQM suffers from a performance degradation in some scenarios (as will be discussed in Section \ref{sec:sim}).

In \cite{zhai17}, an adaptive codebook design and assignment is proposed for SCMA systems that aims to reduce the energy consumption caused by the detection process. The proposed mother constellation has a similar structure as the CQAM constellation. The two-dimensional user-specific rotations are then applied to differentiate the users. As mentioned earlier in Section \ref{sec:constdesign}, those user-specific rotations become questionable in uplink scenarios.
\begin{figure*}
\centering     

\subfigure[]{\label{fig:A165RE1}\includegraphics[width= 0.45\textwidth]{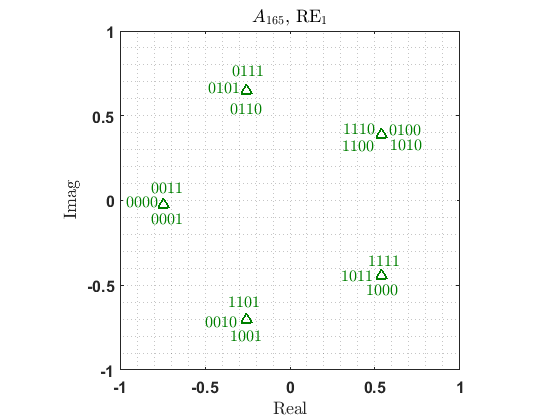}}
\subfigure[]{\label{fig:A165RE2}\includegraphics[width= 0.45\textwidth]{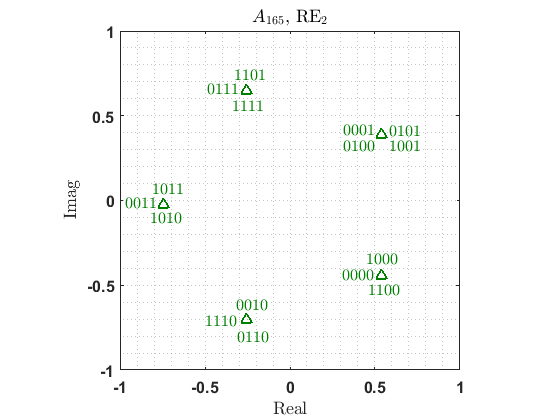}}
\subfigure[]{\label{fig:16bekoRE1}\includegraphics[width= 0.45\textwidth]{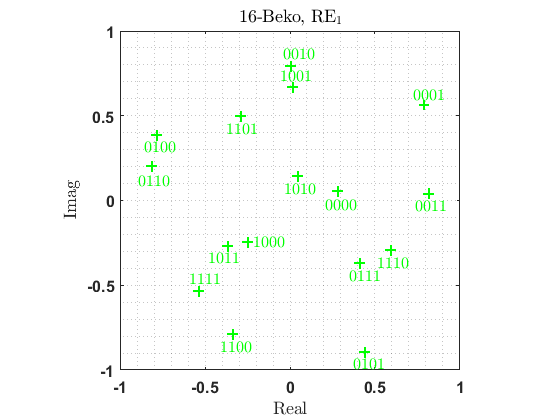}}
\subfigure[]{\label{fig:16bekoRE2}\includegraphics[width= 0.45\textwidth]{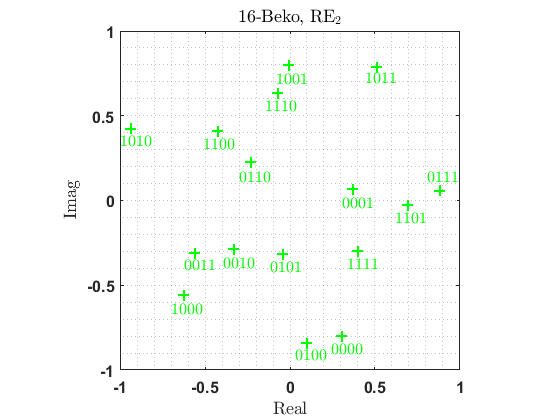}}
\subfigure[]{\label{fig:4pengRE1}\includegraphics[width= 0.45\textwidth]{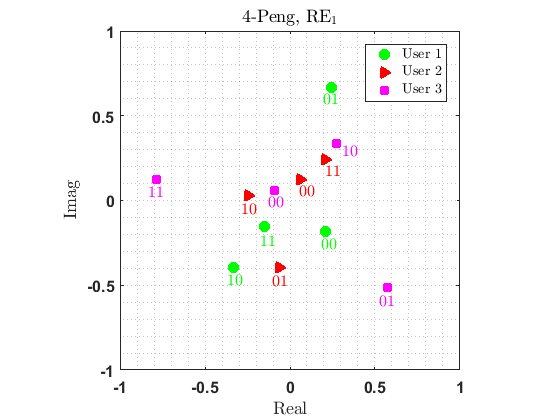}}
\subfigure[]{\label{fig:4pengRE2}\includegraphics[width= 0.45\textwidth]{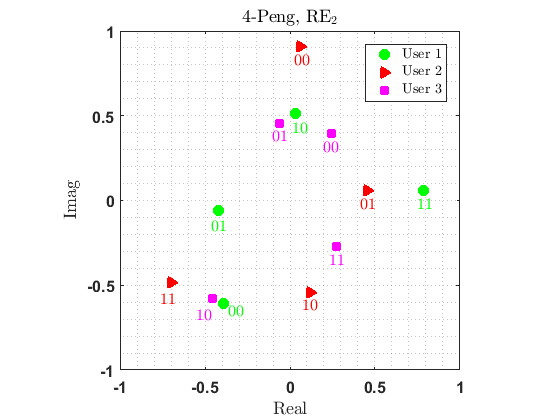}}
\caption{The projection of 4-dimensional constellations over $\textrm{RE}_1$ and $\textrm{RE}_2$: (a) {\color{black}$A_{16,5}$} over $\textrm{RE}_1$, (b) {\color{black}$A_{16,5}$} over $\textrm{RE}_2$, (c) 16-Beko over $\textrm{RE}_1$, (d) 16-Beko over $\textrm{RE}_2$, (e) 4-Peng over $\textrm{RE}_1$, and (f) 4-Peng over $\textrm{RE}_2$.}
\label{fig:const}
\end{figure*}
  \subsubsection{{\color{black}$A_{M,N_d}$}} The $M$-point constellation proposed in \cite{Bao18} with a desired $N_d$ distinct points, and is referred as {\color{black}$A_{M,N_d}$}. Based on the extrinsic information transfer (EXIT) chart, it is shown in \cite{Bao18} that the impact of the constellation and labeling on a single-user system is consistent with that on a multiuser system. In other words, a good constellation and labeling for a single-user system is also good for the multiuser case. A multistage optimization for designing a $2d_v$-dimensional constellation with a low number of projections is then proposed that consists of the following main steps: 1) For a desired $N_d$, a one-dimensional constellation, $\mathcal{A}$, with a good average mutual information (AMI), is used as a base constellation. The $N_d$ amplitude phase-shift keying ($N_d$-APSK) constellation can be constructed using the general APSK procedure in \cite{Hanzo13}, and provides a good AMI. Hence, it is used as the base constellation. 2) Based on the one-dimensional $\mathcal{A}$, a $2d_v$-dimensional constellation is constructed using a permutation function that is provided in \cite{Bao18}. 3) An appropriate labeling based on the EXIT chart is then optimized for the resultant $2d_v$-dimensional constellation.

As an example, we show the projection of the 16-point 4-dimensional constellation proposed in \cite{Bao18} with $N_d=5$, i.e., {\color{black}$A_{16,5}$}, over $\textrm{RE}_1$ and $\textrm{RE}_2$ in Fig.~\ref{fig:A165RE1} and Fig.~\ref{fig:A165RE2}, respectively.

In \cite{He17ldpcscma}, an $M$-point constellation with low number of projections is proposed for non-binary low density parity check (LDPC) coded SCMA that aims to maximize the minimum Euclidian distance, $d_{E,\:\min}$, between the constellation points. As mentioned in Section \ref{sec:kpi}, maximizing $d_{E,\:\min}$ is not a KPI in fading scenarios with independent fading over the $d_v$ REs.
 \subsubsection{$M$-Beko}\label{const:Mbeko}
The $M$-point constellation proposed in \cite{Beko12} for uncoded systems over AWGN channels, by minimizing the average symbol energy for a given $d_{E,\:\min}$ between constellation points. The design of this constellation is formulated as a non-convex optimization problem that is tackled by solving a sequence of convex optimization problems. The indicated constellation is later used in the context of SCMA systems in \cite{nikopour2016systems}. We name the $M$-point constellation proposed in \cite{Beko12} as $M$-Beko\footnote{This constellation is available for different dimensions at the posted URL in \cite{Beko12}.}.

We depict the projection of 4-Beko onto two dimensions with green plus signs over $\textrm{RE}_1$ and $\textrm{RE}_2$ in Fig. \ref{fig:4beko}. As an example, to transmit 11 the user sends $( 0.7543, 0.3852)$ and $(-0.3993,  0.3509)$ over  $\textrm{RE}_1$ and $\textrm{RE}_2$, respectively. In a similar vein, we show projection of 16-Beko onto two dimensions with green plus signs over $\textrm{RE}_1$ and $\textrm{RE}_2$ in Fig. \ref{fig:16bekoRE1} and Fig. \ref{fig:16bekoRE2}, respectively. As we see in Fig. \ref{fig:4beko} and Fig.~\ref{fig:16bekoRE1}--\ref{fig:16bekoRE2}, $M$-Beko is an irregular constellation that does not fall on a grid. This can result in a more complex detection process.
\begin{figure*}
\centering     
\subfigure[]{\label{fig:4psmaRE1}\includegraphics[width= 0.45\textwidth]{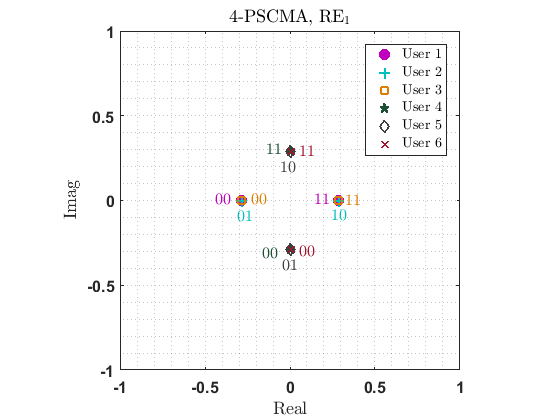}}
\subfigure[]{\label{fig:4pscmaRE2}\includegraphics[width= 0.45\textwidth]{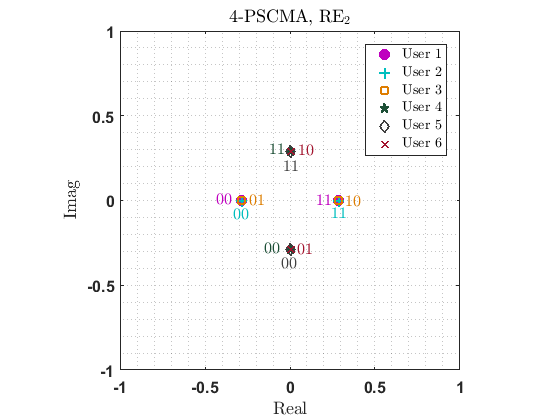}}
\subfigure[]{\label{fig:4pscmaRE3}\includegraphics[width= 0.45\textwidth]{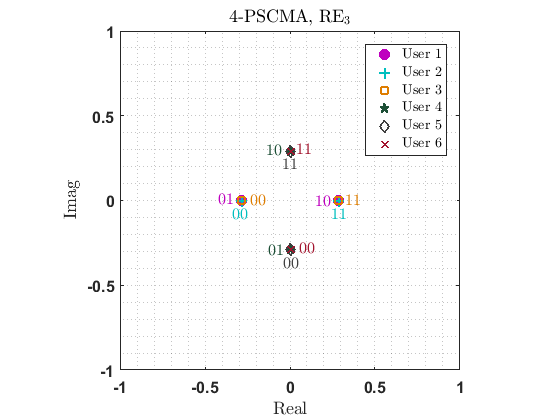}}
\subfigure[]{\label{fig:4pscmaRE4}\includegraphics[width= 0.45\textwidth]{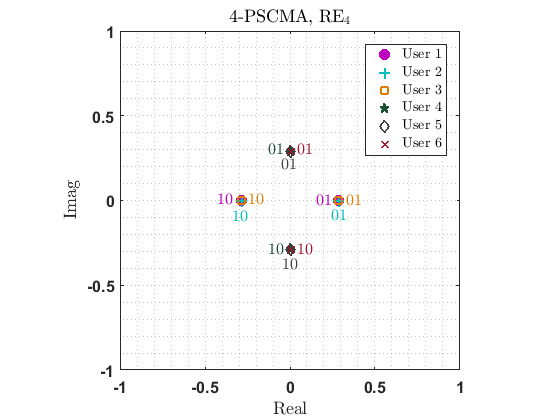}}
\caption{The projection of 4-PSCMA over (a) $\textrm{RE}_1$, (b) $\textrm{RE}_2$, (c) $\textrm{RE}_3$, and (d) $\textrm{RE}_4$.}
\label{fig:4pscma}
\end{figure*}
\subsubsection{$M$-Peng}\label{const:Peng}
An $M$-point codebook that is proposed for uncoded SCMA systems over AWGN channels in \cite{peng17}, and we refer to it as $M$-Peng. Unlike most of the constellations proposed so far in the context of SCMA systems, $M$-Peng is not based on a mother constellation that is common to all users, {\color{black}but instead} is designed by maximizing $d_{E,\:\min}$ between constellation points of \emph{all} users for a power constraint. The design of this constellation is formulated as a non-convex optimization problem that is tackled by using the semidefinite relaxation (SDR)\cite{luo10SDR} technique.

For a scenario with $K=3$ users and $d_v=2$, we depict the 4-Peng constellation over $\textrm{RE}_1$ and $\textrm{RE}_2$ in Fig.~\ref{fig:4pengRE1} and Fig.~\ref{fig:4pengRE2}, respectively. As we see in Fig.~\ref{fig:4pengRE1}--\ref{fig:4pengRE2}, similar to $M$-Beko, $M$-Peng is an irregular constellation, which can lead to a more complex detection process.

In \cite{Klimentyevcodebook17}, an $M$-point constellation is proposed using a genetic algorithm for uncoded SCMA systems over AWGN channels. The constellation in \cite{Klimentyevcodebook17} aims to maximize $d_{E,\:\min}$ between the constellation points for a fixed average energy. This constellation results in two pairs of antipodal codewords, that are then rotated by a 2-dimensional user-specific rotations. Once again, the impact of those user-specific rotations will disappear in the uplink fading scenarios. It is also shown in \cite{Klimentyevcodebook17} that the proposed constellation does not perform well in fading scenarios.

\subsubsection{$M$-PSCMA}\label{const:MPSCMA}
The $M$-point permutation-based SCMA that which we name as $M$-PSCMA, is proposed in \cite{Kulhandjian17} for uncoded SCMA systems. A similar idea was originally employed in the context of CDMA and OFDM in \cite{damour05,shi10,moussa12}. Unlike the conventional SCMA systems, each PSCMA user is not assigned to a fixed user-to-RE mapping matrix, $\mathbf{F}_k$; the position of the non-zero elements in the user-to-RE mapping matrix for each PSCMA user is based on the permutation that is a function of the transmitted bit streams. In other words, the codebook of each user changes with the incoming bits, and therefore MPA cannot be used as the detection algorithm. A low complexity detector is then developed based on an iterative decoding algorithm in \cite{Romano05}.

For a scenario with $K=6$, $N=4$, and $d_v=2$, we depict the 4-PSCMA constellation in Fig.~\ref{fig:4pscma}. All users have an access to the all the available REs. However, they choose 2 out of the 4 REs according to their incoming bit streams. For instance, User 1 employs $\textrm{RE}_1$ and $\textrm{RE}_2$ to send 00 and 11, while it employs $\textrm{RE}_3$ and $\textrm{RE}_4$ to send 01 and 10. User 2 employs $\textrm{RE}_2$ and $\textrm{RE}_3$ to send 00 and 11, while it employs $\textrm{RE}_1$ and $\textrm{RE}_4$ to send 01 and 10. User 3 employs $\textrm{RE}_1$ and $\textrm{RE}_3$ to send 00 and 11, while it employs $\textrm{RE}_2$ and $\textrm{RE}_4$ to send 01 and 10. The codebook of User 4 is the same as the codebook of User 1 except it is rotated by a phase shift of $\pi/2$. Similarly, the codebooks of User 5 and User 6 are the same as the codebook of User 2 and User 3, respectively, but after a rotation of $\pi/2$. As mentioned before, the $\pi/2$ phase shift loses its importance in the uplink scenarios.

\section{Performance Evaluation}\label{sec:sim}
In this section, we first provide the KPIs (Section \ref{sec:kpi}) of an important subset\footnote{{\color{black}In our simulations, we did not assess the performance of the $M$-Peng, $A_{N_m,N_d}$, and $M$-PSCMA constellations due to the following reasons: The design of the $M$-Peng constellation depends on $K$ and $N$ \cite{peng17}, and we did not have access to this constellation designed for our system parameters. The $A_{N_m,N_d}$ constellation is designed specifically for BICM with iterative decoding and detection \cite{Bao18}. However, we use non-iterative detection and decoding in our simulations. Also, the $M$-PSCMA constellation requires a special receiver \cite{Kulhandjian17}, and to maintain focus, we only use MPA with non-iterative detection and decoding.}} of the constellations described in Section \ref{sec:constellations}, namely, $M$-LDS (Section \ref{const:MLDS}), T-$M$QAM (Section \ref{const:TMqam}), $M$LQAM (Section \ref{const:LMqam}), $M$-Bao (Section \ref{const:Mbao}), 16HQAM (Section \ref{const:16Hqam}), $M$CQAM (Section \ref{const:MCqam}), and $M$-Beko (Section \ref{const:Mbeko}). We then evaluate the performance of those constellations over the different channel models presented in Section \ref{sec:chnmodel} through extensive MATLAB simulations. For convenience, in our all performance evaluation curves, we stick with the same color/line specification as in Fig.~\ref{fig:16const1}--Fig.~\ref{fig:16const2} and Fig.~\ref{fig:16bekoRE1}--\ref{fig:16bekoRE2}. That is, the $M$-LDS constellation is represented by magenta left-pointing triangles,  T$M$QAM by red circles, $M$LQAM by blue asterisks, $M$-Bao by cyan squares, 16HQAM by brown crosses, $M$CQAM by black points, and $M$-Beko by green plus signs.
\subsection{Constellations from {\color{black}the Standpoint of KPIs}}\label{subsec:kpieval}
As discussed in Section \ref{sec:kpi}, the KPIs of $M$-point $2d_v$-dimensional constellations over different channel scenarios under study are as follows: The minimum Euclidian distance, $d_{E,\:\min}$, the Euclidian kissing number, $\tau_E$, the minimum product distance, $d_{P,\:\min}$, the product kissing number, $\tau_P$, the modulation diversity order, $L$, the number of distinct points, $N_d$, and whether the constellation is Gray-labeled or not. 
We provide those properties of the 4-point and 16-point 4-dimensional constellations under study in Table \ref{table:cmplxtm4ary} and Table \ref{table:cmplxtm16ary}, respectively.
\begin{table}[!t]
\caption{{\color{black}KPIs of 4-point constellations}}
\centering 
\scalebox{0.8}{
\begin{tabular}{cccccccc } 
\multicolumn{8}{c}{} \\ [0.7ex]
 \hline 
  & $d^2_{E,\min}$ &$\tau_E$&$d_{P,\min}^2$&$\tau_P$&$L$&${\color{black}N_d}$&Gray-labeled\\ 
 \hline
 T4QAM &2&2&$0.64$&2& 2&4&Yes \\
 4LQAM&2&2& 2&2&1&2&Yes\\
 4CQAM &2&2& 1&2&1&3&Yes\\
  4-Beko&$\cong2.67$&3&$\cong 0.29$&$1/2$&2&4&No\\
  4-Bao &2&2&$\cong1$&2&2&4&Yes\\
  4-LDS&2&2&1&2&2&4&Yes\\
 \hline 
\end{tabular}
}
\label{table:cmplxtm4ary} 
\end{table}
\begin{table}[!t]
\caption{{\color{black}KPIs of 16-point constellations}}
\centering 
\scalebox{0.8}{
\begin{tabular}{cccccccc }
\multicolumn{8}{c}{} \\ [0.5ex]
 \hline 
  &$d_{E,\min}^2$ &$\tau_E$&$d_{P,\min}^2$& $\tau_P$&$L$&${\color{black}N_d}$& Gray-labeled\\ 
 \hline
 T16QAM &1&4&0.16&4&2&16&Yes  \\
 16LQAM&1&4&$0.25$&4&1&9& Yes\\
 16HQAM&1&4&1&8&1&4& Yes \\
 16CQAM & $\cong 1.03$&2&$\cong0.21$ &2&1&9&No \\
 16-Beko&$\cong 1.30$&{\color{black}7.75}&$\cong 0.02$&1/8&2&16&No\\
 16-Bao &1&4&0.25&4&2&{\color{black}16}&Yes \\
 16-LDS&$0.4$&3&0.04&3&2&16&Yes \\
 \hline 
\end{tabular}}
\label{table:cmplxtm16ary} 
\end{table}
From Table \ref{table:cmplxtm4ary} and Table \ref{table:cmplxtm16ary}, we observe the following:
\begin{itemize}
  \item 4-Beko and 16-Beko have the highest $d_{E,\:\min}$ compared with the other 4-point and 16-point constellations. This is due to the fact that these constellations are proposed for AWGN channels, wherein $d_{E,\:\min}$ is a KPI, and has been explicitly considered in the design process \cite{Beko12}. This notwithstanding, both 4-Beko and 16-Beko also have the highest $\tau_E$ compared with other 4-point and 16-point constellations, which as we will discuss later, results in a performance degradation in some scenarios.  
  \item From Section \ref{sec:constellations}, there are number of overlaps among the projections of the constellation points in $M$LQAM, 16HQAM, and $M$CQAM. As such, the modulation diversity order of $M$LQAM, 16HQAM, and $M$CQAM is $L=1$. {\color{black}Moreover, from \eqref{Nd}, $N_d$ of the constellations are attained.}
  \item As mentioned in Section \ref{const:Mbao}, $M$-Bao is proposed for FIC scenarios. As $d_{P,\: \min}$ is a KPI in FIC, both 4-Bao and 16-Bao have the highest $d_{P,\: \min}$ compared with other constellations with $L=2$.
  \item Due to the structure of $M$-Beko, it is not possible to use Gray labeling. Also, 16$C$QAM was not Gray-labeled when proposed in \cite{cqam17}.
\end{itemize}

In what follows, we evaluate the performance of different constellations introduced in Section \ref{sec:constellations} under different channel scenarios described in Section \ref{sec:chnmodelKPI}. Since the 2-dimensional user-specific rotations lose their importance in uplink scenarios, we assume all users employ the same constellations as the mother constellations provided in Section \ref{sec:constellations}. We set $K=6$, $N=4$, $d_v=2$, $d_c=3$,\footnote{Note that this has been the most common set up so far in different contexts of SCMA systems e.g., \cite{scma,taherzadeh14,Baoconstellation,Bao16sphericalcode,cqam17,vameghestahbanati2017polar,meSD1CL17, Sergienko16,Dai16_polar,Dai17TVT,Dabiri18}.} with a single antenna at all transmitters, and a single antenna at the receiver. We also use {\color{black}the user-to-RE indicator matrix provided in \eqref{S}, set} 3 MPA iterations for the 4-point constellations, and 5 MPA iterations for the 16-point constellations \cite{meSD1CL17}. 
\subsection{Uncoded Scenarios}\label{subsec:uc}
In this section, we assess the SER and the BER performance of 4-point and 16-point 4-dimensional constellations with respect to SNR, {\color{black}that is defined as in \eqref{snruncoded},} over FSC (Case \ref{case:fsc}) and FIC (Case \ref{case:fdc}). 
We define SNR as the average energy per bit of the constellation divided by the noise variance. That is, 
\begin{eqnarray}\label{snruncoded}
  \textrm{SNR} &=& \frac{E_s}{L_M\:N_0} \nonumber \\
               &=& \frac{E_b}{N_0},
\end{eqnarray}
where $E_b$ denotes the average energy per {\color{black}uncoded} bit of the constellation{\color{black}, as $E_s$ is given by \eqref{eqn:Es}}.

\subsubsection{SER performance}\label{subsec:ser}
\begin{figure*}[t]
\centering     
\subfigure[]{\label{fig:FSC4ary}\includegraphics[width= 0.4\textwidth]{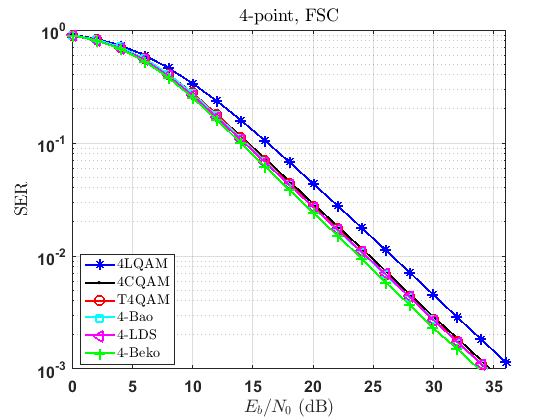}}
\subfigure[]{\label{fig:FSC16ary}\includegraphics[width= 0.4\textwidth]{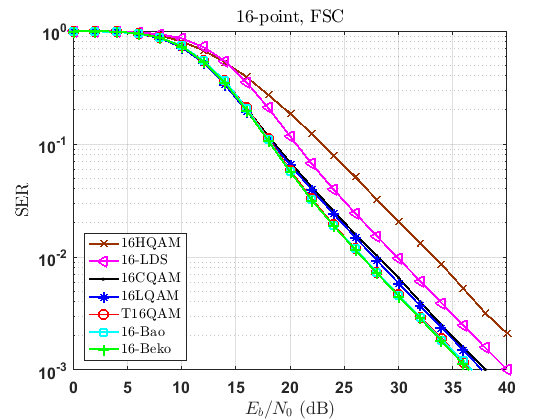}}
\subfigure[]{\label{fig:FIC4ary}\includegraphics[width= 0.4\textwidth]{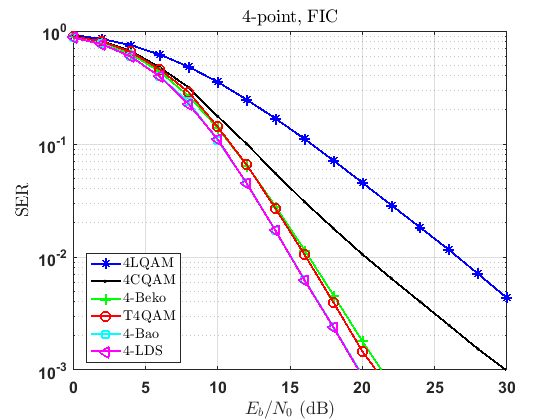}}
\subfigure[]{\label{fig:FIC16ary}\includegraphics[width= 0.4\textwidth]{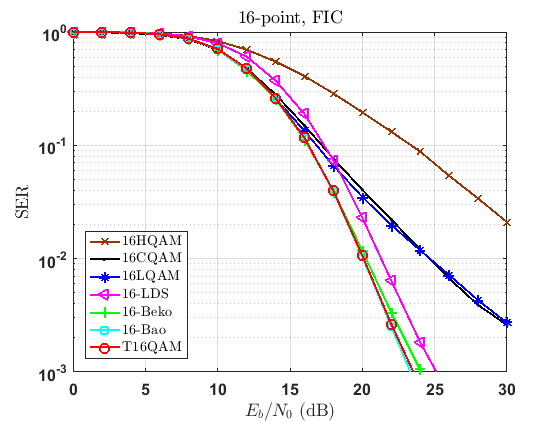}}
\caption{SER performance of uncoded SCMA systems with (a) 4-point constellations over FSC (Case \ref{case:fsc}), (b) 16-point constellations over FSC {\color{black}(Case \ref{case:fsc})}, (c) 4-point constellations over FIC (Case \ref{case:fdc}), and (d) 16-point constellations over FIC {\color{black}(Case \ref{case:fdc})}.}
\end{figure*}
In Fig.~\ref{fig:FSC4ary}--\ref{fig:FSC16ary}, we compare the SER performance of uncoded SCMA systems with different 4-point and 16-point constellations over FSC channels (Case \ref{case:fsc}). 
 As discussed in Section \ref{sec:kpi}, {\color{black}the number of distinct points, $N_d$}, the minimum Euclidian distance, $d_{E, \min}$, and the Euclidian kissing number, $\tau_E$, are the KPIs for the FSC scenario. {\color{black}Since in FSC, each user observes the \emph{same} channel coefficients over the $d_v=2$ REs, we expect that the channel diversity order to be 1, i.e., a change of 10 dB in $E_b/N_0$ per a change of one decade in SER \cite{Rappaport}.

As shown in Fig.~\ref{fig:FSC4ary}, we note that amongst the 4-point constellations, the 4-Beko constellation (Section \ref{const:Mbeko}), which is carefully designed for the AWGN in \cite{Beko12}, performs the best in this scenario.  
As from Table \ref{table:cmplxtm4ary}, 4-Beko (Section \ref{const:Mbeko}) has {\color{black}$N_d=4$}, the highest $d_{E, \min}$, and a comparable $\tau_E$ with other 4-point constellations. 
Moreover, from Table \ref{table:cmplxtm4ary}, since 4-Bao (Section \ref{const:Mbao}), 4-LDS (Section \ref{const:MLDS}), and T4QAM (Section \ref{const:TMqam}) have {\color{black}$N_d=4$} and the same $d_{E, \min}$ and $\tau_E$, they behave similarly in FSC. However, 4CQAM (Section \ref{const:MCqam}) performs slightly worse due to its {\color{black}$N_d=3$}. We also observe that 4LQAM (Section \ref{const:LMqam}) has the {\color{black}lowest $N_d=2$}, and thus falls behind the other constellations in this scenario.  

Amidst the 16-point constellations, we notice that 16-Beko (Section \ref{const:Mbeko}), T16QAM (Section \ref{const:TMqam}), and 16-Bao (Section \ref{const:Mbao}) outperform the other constellations in Fig.~\ref{fig:FSC16ary}. Referring to Table \ref{table:cmplxtm16ary}, 16-Beko, T16QAM, 16-Bao, and 16-LDS (Section \ref{const:MLDS}) have {\color{black}an $N_d$ of 16}. However, we observe in Fig.~\ref{fig:FSC16ary} that 16-LDS lags behind 16-Beko, T16QAM, and 16-Bao due to its lower $d_{E,\min}$. Further, although 16-Beko has the highest $d_{E, \min}$ compared with the others, it is also with the highest $\tau_E$. As such, it performs similar to T16QAM and 16-Bao (which have the same $d_{E, \min}$ and $\tau_E$). We also note that 16HQAM (Section \ref{const:16Hqam}) has the {\color{black}lowest $N_d=4$}, and thus it stays behind all the other constellations in this scenario.

We depict the SER performance of uncoded SCMA systems with different 4-point and 16-point constellations over FIC channels (Case \ref{case:fdc}) in Fig.~\ref{fig:FIC4ary} and Fig.~\ref{fig:FIC16ary}, respectively. 
 As addressed in Section \ref{sec:kpi}, {\color{black}$N_d$}, the modulation diversity order, $L$, the minimum product distance, $d_{P, \min}$, and the product kissing number, $\tau_P$, are the KPIs for the FIC scenario. {\color{black}Since in FIC, each user observes \emph{different} channel coefficients over the $d_v=2$ REs, the channel diversity order is 2, i.e., a change of 5 dB in $E_b/N_0$ per a change of one decade in SER.} 

Fig.~\ref{fig:FIC4ary} shows the 4-point constellations in the FIC scenario. From Table \ref{table:cmplxtm4ary}, 4-Bao, 4-LDS, T4QAM, and 4-Beko have $L=2$ and {\color{black}$N_d=4$}, and so perform similarly. Among these constellations, 4-Bao and 4-LDS have the highest $d_{P,\min}$ and the same $\tau_P$, and as such, they outperform the others. Moreover, {\color{black}compared with the constellations with $L=2$, the performance of 4-Beko is degraded by its lowest $d_{P, \min}$.} 
Also, since 4LQAM has the {\color{black}lowest $N_d$} and $L$ compared with the others, it lags behind them.

As Fig.~\ref{fig:FIC16ary} presents, amidst the 16-point constellations, 16-Bao beats the others over the FIC scenario. Referring to Table \ref{table:cmplxtm16ary}, 16-Bao, T16QAM, 16-Beko, and 16-LDS have $L=2$ and {\color{black}$N_d=16$}. Among these constellations, we see that 16-Bao has the highest $d_{P,\min}$ and the same $\tau_P$ as T16QAM and 16-LDS. As such, it outperforms the others. Note that although 16-LDS has a slightly higher $d_{P,\min}$ compared with 16-Beko, it still stays behind it due to its higher $\tau_P$. Further, 16HQAM performs very poorly in this scenario due its low $L$ and {\color{black}$N_d$}. 

\subsubsection{BER performance}\label{subsec:ber}\begin{figure*}[t]
\centering     
\subfigure[]{\label{fig:FSC4aryber}\includegraphics[width= 0.4\textwidth]{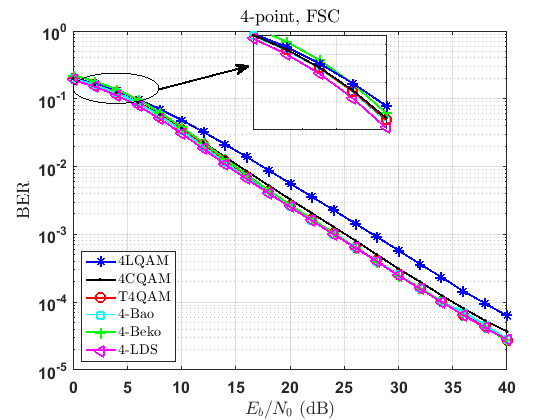}}
\subfigure[]{\label{fig:FSC16aryber}\includegraphics[width= 0.4\textwidth]{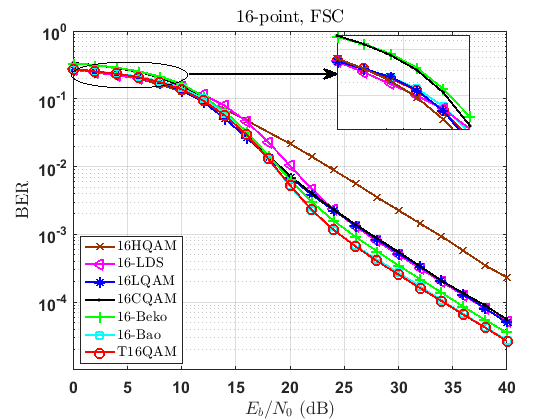}}
\subfigure[]{\label{fig:FIC4aryber}\includegraphics[width= 0.4\textwidth]{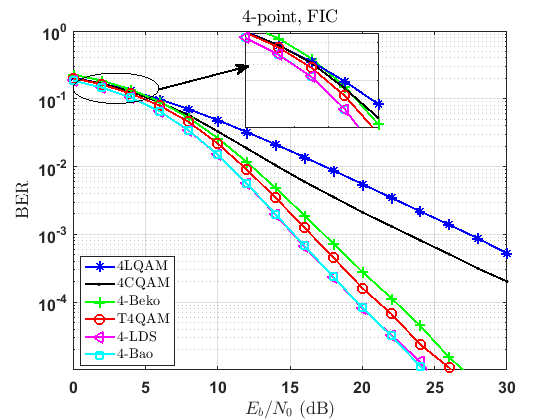}}
\subfigure[]{\label{fig:FIC16aryber}\includegraphics[width= 0.4\textwidth]{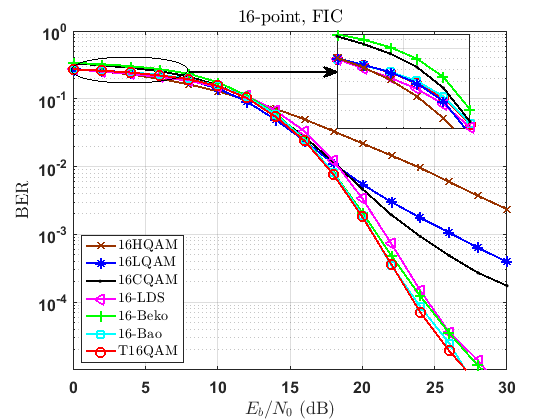}}
\caption{BER performance of uncoded SCMA systems with (a) 4-point constellations over FSC (Case \ref{case:fsc}), (b) 16-point constellations over FSC, (c) 4-point constellations over FIC (Case \ref{case:fdc}), and (d) 16-point constellations over FIC.}
\end{figure*}
We compare the BER performance of uncoded SCMA systems introduced in Section \ref{subsec:ser}, in Fig.~\ref{fig:FSC4aryber}--\ref{fig:FIC16aryber}. As mentioned in Section \ref{sec:kpi}, in addition to the KPIs for the SER performance of FSC and FIC scenarios mentioned above, bit-labeling has a significant impact on the BER performance of uncoded systems. Moreover, the behavior of the multi-user detector at different SNRs affects the performance of the system. For the illustrative purpose,
the BER performance of the uncoded systems at low SNRs is magnified in Fig.~\ref{fig:FSC4aryber}--\ref{fig:FIC16aryber}. We note that the presence of multi-user interference (MUI) changes the ordering of the BER curves at low SNRs.

From Table \ref{table:cmplxtm4ary} and Table \ref{table:cmplxtm16ary}, $M$-Beko and 16CQAM
constellations are not Gray-labeled. Therefore, we observe in Fig.~\ref{fig:FSC4aryber}--\ref{fig:FIC16aryber} that compared with their SER performance, $M$-Beko and 16CQAM perform differently with respect to other constellations. For instance, we see in  Fig.~\ref{fig:FSC4aryber} and Fig.~\ref{fig:FSC16aryber} that unlike their SER performance in Fig.~\ref{fig:FSC4ary} and Fig.~\ref{fig:FSC16ary}, 4-Beko and 16-Beko do not outperform the other constellations. 

\subsection{High-Rate Turbo-coded Scenarios}\label{subsec:highratetc}
\begin{figure*}[t]
\centering     
\subfigure[]{\label{fig:FFSChigh4ary}\includegraphics[width= .4\textwidth]{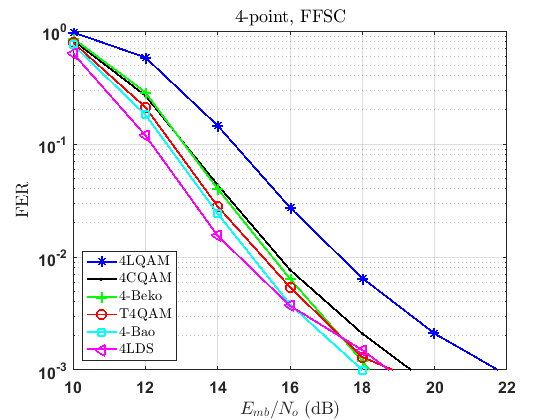}}
\subfigure[]{\label{fig:FFSChigh16ary}\includegraphics[width= 0.4\textwidth]{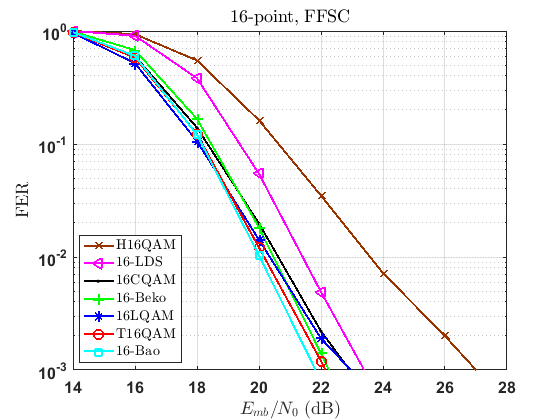}}
\subfigure[]{\label{fig:FFDChigh4ary}\includegraphics[width= 0.4\textwidth]{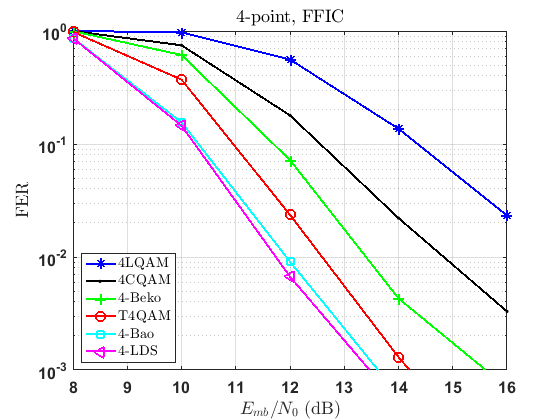}}
\subfigure[]{\label{fig:FFDChigh16ary}\includegraphics[width= 0.4\textwidth]{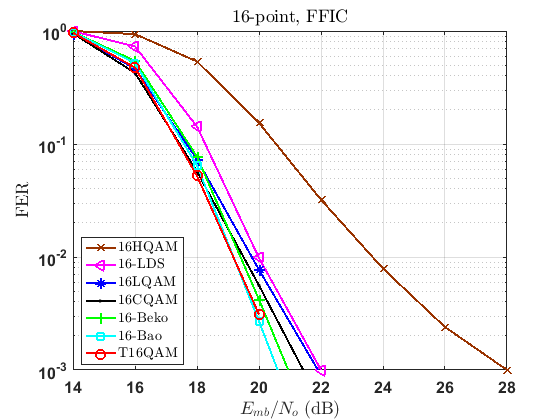}}
\caption{FER performance of $R=5/6$ turbo-coded SCMA systems with (a) 4-point constellations over FFSC (Case \ref{case:ffsc}), (b) 16-point constellations over FFSC, (c) 4-point constellations over FFIC (Case \ref{case:ffdc}), and (d) 16-point constellations over FFIC.}
\end{figure*}
In this section, we evaluate the frame error rate (FER) performance of high-rate turbo-coded SCMA systems with 4-point and 16-point 4-dimensional constellations with respect to SNR  over FFSC (Case \ref{case:ffsc}), FFIC (Case \ref{case:ffdc}), SFSC (Case \ref{case:sfsc}), and SFIC (Case \ref{case:sfdc}). We define SNR as the average energy per message bit of the constellation divided by the noise variance. That is, 
\begin{eqnarray}\label{snrcoded}
  \textrm{SNR} &=& \frac{E_s}{R\:L_M\:N_0} \nonumber \\
               &=& \frac{E_{mb}}{N_0},
\end{eqnarray}
where $E_{mb}$ denotes the average energy per bit of the constellation, and $R$ is the rate of the turbo code.

We employ an LTE turbo code \cite{3GPP} with a rate of $R=5/6$ and a codeword length of {\color{black} $N_c=120$}. In our simulations, we use 4 turbo iterations. 
{\color{black}We use the turbo encoder followed by the rate-matcher used in the LTE standard in \cite{3GPP}. The LTE rate-matcher includes interleaving between the encoder and the signal mapper and is described in \cite{3GPP}.}
Interested readers can refer to \cite{zarrinkoub2014understanding} for more information about implementing the LTE turbo code at different rates.


 Fig.~\ref{fig:FFSChigh4ary}--\ref{fig:FFSChigh16ary} show the FER performance of turbo-coded SCMA systems with $R=5/6$ and different 4-point and 16-point constellations for the FFSC scenario. As discussed in Section \ref{sec:kpi}, the type of bit-labeling, the SNR operating region, {\color{black}$N_d$}, $d_{E, \min}$, and $\tau_E$ are the KPIs for this scenario. 

Amongst the 4-point constellations, we note that 4-LDS is the best choice in Fig.~\ref{fig:FFSChigh4ary}.
It is worthwhile to note that 4-Bao lags behind 4-LDS in this scenario. This can be justified through the BER performance of uncoded systems at {\color{black}mid-range} SNRs in Fig.~\ref{fig:FSC4aryber}.\footnote{{\color{black}Of note, the energy per message bit in Fig.~\ref{fig:FFSChigh4ary} and the energy per uncoded bit in Fig.~\ref{fig:FSC4aryber} are related by $10\log\left(R\right)$.}}
Moreover, similar to the BER performance in uncoded system in Fig.~\ref{fig:FSC4aryber}, although from Table \ref{table:cmplxtm4ary}, 4-Beko has {\color{black}$N_d=4$}, the highest $d_{E, \min}$, and a comparable $\tau_E$ with the other 4-point constellations, it does not perform as well as 4-LDS due to its bit-labeling. We also observe that 4LQAM has the {\color{black}lowest $N_d=2$}, and thus falls behind the others.  
On the other hand, amidst the 16-point constellations, we notice that 16-Bao and T16QAM outperform the others in Fig.~\ref{fig:FFSChigh16ary}. Referring to Table \ref{table:cmplxtm16ary}, 16-Beko, T16QAM, 16-Bao, and 16-LDS  have {\color{black}$N_d=16$}. However, we observe in Fig.~\ref{fig:FFSChigh16ary} that 16-LDS lags behind T16QAM, 16-Bao, 16-Beko due to its comparable $\tau_E$, and its lowest $d_{E,\min}$. Further, although 16-Beko has the highest $d_{E, \min}$, it also has the highest $\tau_E$, and is not Gray-labeled. As such, it lags behind 16-Bao and T16QAM. We also note that similar to the FSC case (Fig.~\ref{fig:FSC16ary}), 16HQAM has the {\color{black}lowest $N_d=4$}, and thus it stays behind all the other constellations in this scenario.

We depict the FER performance of turbo-coded SCMA systems with $R=5/6$ and different 4-point and 16-point constellations over FFIC channels in Fig.~\ref{fig:FFDChigh4ary} and Fig.~\ref{fig:FFDChigh16ary}, respectively. 
 As addressed in Section \ref{sec:kpi}, the type of bit-labeling, the SNR operating region, $N_d$, $L$, $d_{p, \min}$, and $\tau_p$ are the KPIs for the FFIC scenario. 

As Fig.~\ref{fig:FFDChigh4ary} shows, amongst the 4-point constellations, 4-LDS and 4-Bao outperform the others in the FFIC scenario. Following the similar illustration as in the FIC case (Fig.~\ref{fig:FIC4ary}) and from Table \ref{table:cmplxtm4ary}, 4-Bao, 4-LDS, T4QAM, and 4-Beko have the highest $L$ with {\color{black}$N_d=4$}. Among these constellations, 4-Bao and 4-LDS have the highest $d_{P,\min}$ and the same $\tau_P$. As such, both 4-Bao and 4-LDS outperform the others. Moreover, compared with the other constellations with $L=2$, the performance of 4-Beko is deteriorated by its lowest $d_{P, \min}$ and its bit-labeling. 
Also, similar to the FIC case, since 4LQAM has the {\color{black}lowest $N_d$} and $L$, it lags behind the others. As Fig.~\ref{fig:FFDChigh16ary} presents, amidst the 16-point constellations, 16-Bao and T16QAM beat the others over the FFIC scenario. As for the FIC case (Fig.~\ref{fig:FIC16ary}), and by referring to Table \ref{table:cmplxtm16ary}, 16-Bao, T16QAM, 16-Beko, and 16-LDS have the highest $L$ with {\color{black}$N_d=16$}. Among these, we see that 16-Bao has the highest $d_{P,\min}$, a comparable $\tau_P$ with T16QAM and 16-LDS, and a similar bit-labeling. 
Moreover, 16HQAM performs very poorly in this scenario due its lowest $L$ and {\color{black}lowest $N_d$}. 
\begin{figure*}[t]
\centering     
\subfigure[]{\label{fig:SFSChigh4ary}\includegraphics[width= .4\textwidth]{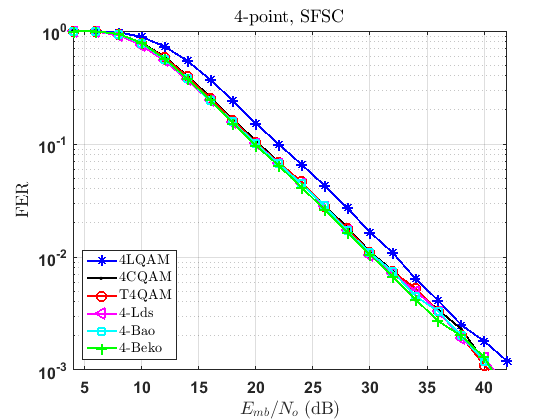}}
\subfigure[]{\label{fig:SFSChigh16ary}\includegraphics[width= 0.4\textwidth]{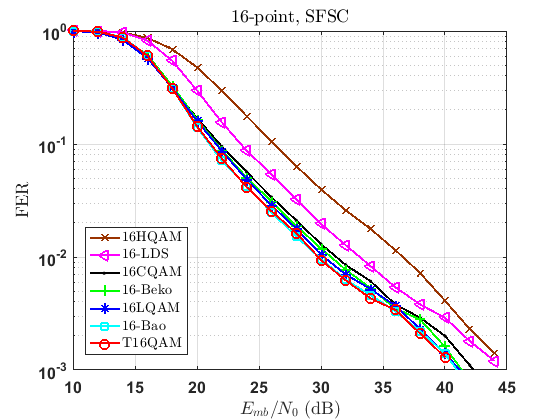}}
\subfigure[]{\label{fig:SFDChigh4ary}\includegraphics[width= 0.4\textwidth]{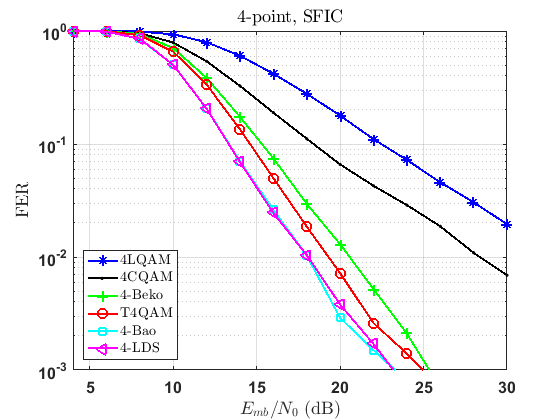}}
\subfigure[]{\label{fig:SFDChigh16ary}\includegraphics[width= 0.4\textwidth]{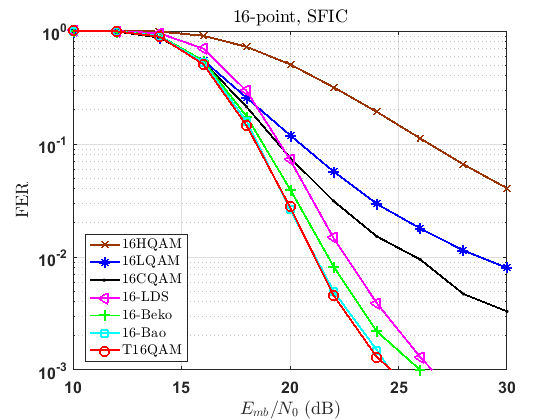}}
\caption{FER performance of $R=5/6$ turbo-coded SCMA systems with (a) 4-point constellations over SFSC (Case \ref{case:sfsc}), (b) 16-point constellations over SFSC, (c) 4-point constellations over SFIC (Case \ref{case:sfdc}), and (d) 16-point constellations over SFIC.}
\end{figure*}

In Fig.~\ref{fig:SFSChigh4ary}--\ref{fig:SFDChigh16ary}, we compare the FER performance of the systems described above over SFSC and SFIC channels. {\color{black}Note that since in slow fading scenarios each channel coefficient is constant for the duration of transmission of the whole codeword, the system performs poorly. Hence, in slow fading scenarios, higher SNR regions are mainly of interest.}

Similar to the FFSC case, the type of bit-labeling, the SNR operating region, {\color{black}$N_d$}, $d_{E, \min}$, and $\tau_E$ are the KPIs in SFSC. Also, similar to the FFIC case, the type of bit-labeling, the SNR operating region, {\color{black}$N_d$}, $L$, $d_{P, \min}$, and $\tau_P$ are the KPIs in SFIC.

From Fig.~\ref{fig:SFSChigh4ary} and Fig.~\ref{fig:SFSChigh16ary}, we see that the trends of different 4-point and 16-point constellations in SFSC are consistent with the trends of the BER performance of uncoded systems at high SNRs in the FSC case, i.e., Fig.~\ref{fig:FSC4aryber} and Fig.~\ref{fig:FSC16aryber}. Further, from Fig.~\ref{fig:SFDChigh4ary} and Fig.~\ref{fig:SFDChigh16ary}, we observe that the trends of different 4-point and 16-point constellations in SFIC {\color{black}for a target FER of $10^{-2}$ are consistent with the trends of the BER performance of uncoded systems at their corresponding SNRs ($\cong17.2$ dB and $\cong20.2$ dB in the 4-point and 16-point cases, respectively)} in the FIC case, i.e., Fig.~\ref{fig:FIC4aryber} and Fig.~\ref{fig:FIC16aryber}.



\subsection{Low-Rate Turbo-coded Scenarios}\label{subsec:lowratetc}
\begin{figure*}[t]
\centering     
\subfigure[]{\label{fig:FFSClow4ary}\includegraphics[width= .4\textwidth]{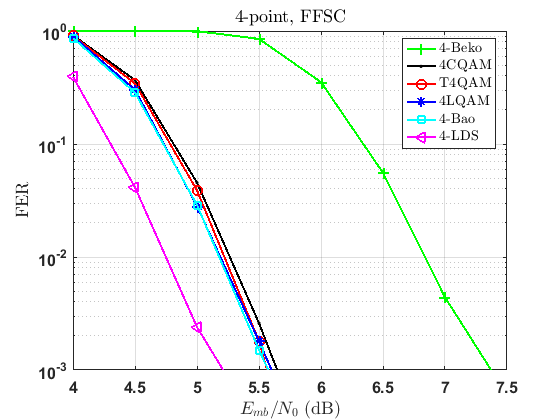}}
\subfigure[]{\label{fig:FFSClow16ary}\includegraphics[width= 0.4\textwidth]{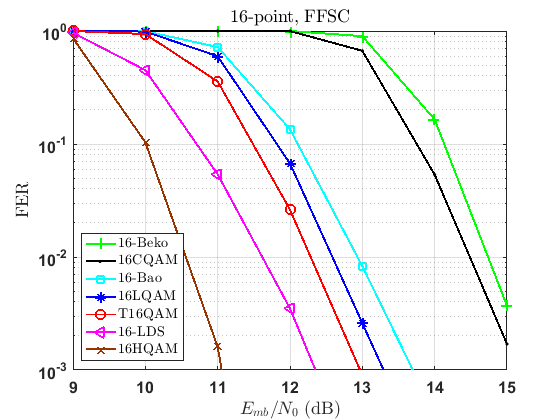}}
\subfigure[]{\label{fig:FFDClow4ary}\includegraphics[width= 0.4\textwidth]{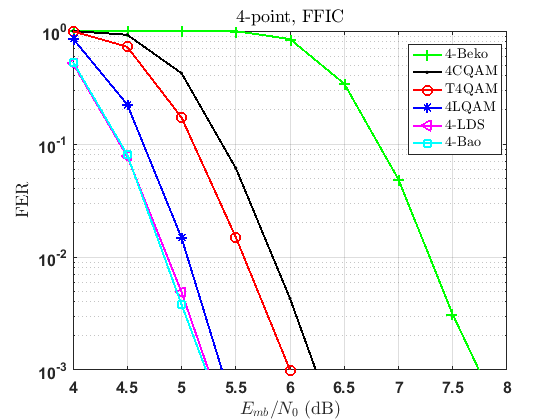}}
\subfigure[]{\label{fig:FFDClow16ary}\includegraphics[width= 0.4\textwidth]{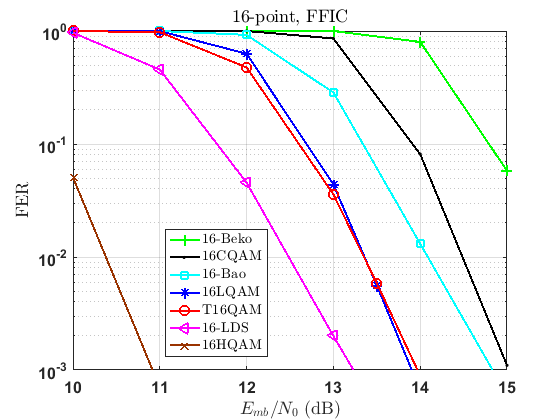}}
\caption{FER performance of $R=1/3$ turbo-coded SCMA systems with (a) 4-point constellations over FFSC (Case \ref{case:ffsc}), (b) 16-point constellations over FFSC, (c) 4-point constellations over FFIC (Case \ref{case:ffdc}), and (d) 16-point constellations over FFIC.}
\end{figure*}
In this section, we evaluate the FER performance of low-rate turbo-coded SCMA systems with 4-point and 16-point 4-dimensional constellations with respect to SNR in different scenarios. We define SNR as the same way as \eqref{snrcoded}. We employ an LTE turbo code with a rate of $R=1/3$, a codeword length of $N_c=2028$, and 4 turbo iterations.

In Fig.~\ref{fig:FFSClow4ary}--\ref{fig:SFDClow16ary}, we compare the FER performance of turbo-coded SCMA systems with $R=1/3$ and different constellations over FFSC, FFIC, SFSC, and SFIC.  

As mentioned earlier, depending on the behavior of the multi-user detector, the system performs differently at {\color{black}different} SNRs.  {\color{black}In fast fading {\color{black}and especially in the presence of low-rate (more powerful) code}, due to the high amount of diversity introduced by the channel, a low FER can be achieved at lower SNRs.} Fig.~\ref{fig:FFSClow4ary}--\ref{fig:FFDClow16ary} evaluate the performance of $R=1/3$ systems at FFSC and FFIC. We see that the trends of different constellations are fairly consistent with the trends of BER performance of the uncoded systems at {\color{black}their corresponding} low SNRs in Fig.~\ref{fig:FSC4aryber}--\ref{fig:FIC16aryber}. More specifically, 4-LDS outperforms the other 4-point constellations in FFSC and FFIC, while 16HQAM performs better than the other 16-point constellations in FFSC and FFIC. 

In Fig.~\ref{fig:SFSClow4ary}--\ref{fig:SFSClow16ary}, we show the FER performance of the systems described above with $R=1/3$ over SFSC. We note that unlike the case of  high-rate codes (Section \ref{subsec:highratetc}), in the presence of a low rate code, the effect of {\color{black}$N_d$} is less significant in this scenario. From Fig.~\ref{fig:SFSClow4ary} and referring to Table \ref{table:cmplxtm4ary}, we see that since all the Gray-labeled constellations have the same $d_{E, \min}$ and $\tau$, they perform similarly. On the other hand, 4-Beko lags behind the other constellations due to its bit-labeling. From Fig.~\ref{fig:SFSClow16ary} and referring to Table \ref{table:cmplxtm16ary}, we observe that among the Gray-labeled constellations, T16QAM, 16LQAM, 16HQAM, and 16-Bao, outperform the others. This is because they all have the same $d_{E, \min}$ and $\tau_E$. Further, 16-LDS performs worse than the other Gray-labeled constellations due to its lower $d_{E, \min}$. Also, 16-Beko and 16CQAM are behind the others due to their non-Gray bit-labeling.
\begin{figure*}[t]
\centering     
\subfigure[]{\label{fig:SFSClow4ary}\includegraphics[width= 0.4\textwidth]{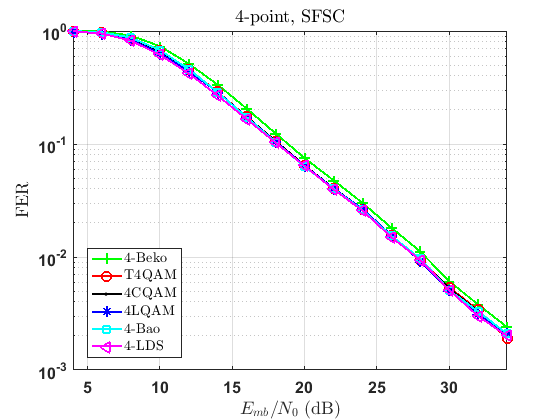}}
\subfigure[]{\label{fig:SFSClow16ary}\includegraphics[width= 0.4\textwidth]{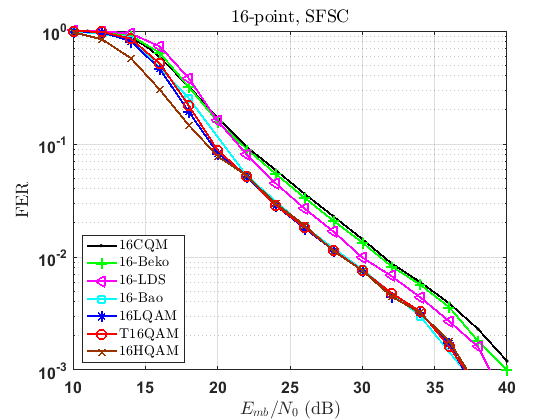}}
\subfigure[]{\label{fig:SFDClow4ary}\includegraphics[width= 0.4\textwidth]{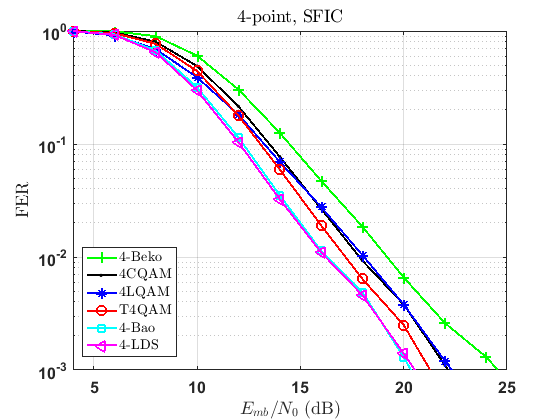}}
\subfigure[]{\label{fig:SFDClow16ary}\includegraphics[width= 0.4\textwidth]{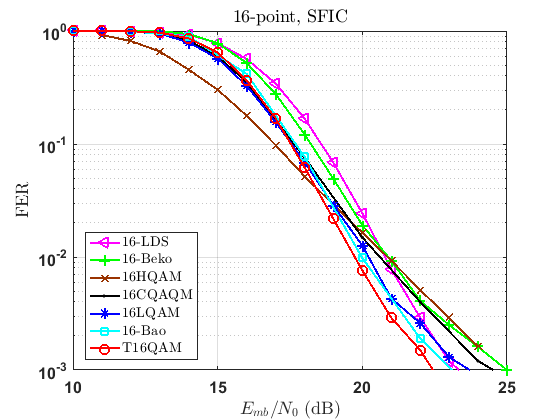}}
\caption{FER performance of $R=1/3$ turbo-coded SCMA systems with (a) 4-point constellations over SFSC (Case \ref{case:sfsc}), (b) 16-point constellations over SFSC, (c) 4-point constellations over SFIC (Case \ref{case:ffdc}), and (d) 16-point constellations over SFIC.}
\end{figure*}

Fig.~\ref{fig:SFDClow4ary}--\ref{fig:SFDClow16ary} present the FER performance of the systems mentioned above over SFIC. We see that from Fig.~\ref{fig:SFDClow4ary} and Table \ref{table:cmplxtm4ary} that 4-Bao and 4-LDS outperform the other constellations, as they both have similar KPI values. Nevertheless, from Fig.~\ref{fig:SFDClow16ary}, we see that T16QAM outperforms the others. {\color{black}As pointed in \cite{taherzadeh14}, T16QAM is constructed similar to \cite{Boutros98}. It was shown by Knopp and Humblet that constellations constructed the same way as \cite{Boutros96,Boutros98} are well suited for block fading channels with independent fading \cite{Knopp2000}.} 
{\color{black}\subsection{Summary of Results}
\begin{table}[!t]
\caption{{\color{black}KPIs of multidimensional constellations for different channel scenarios}}
\centering 
{\color{black}
\scalebox{0.8}{
\begin{tabular}{ccccccc}
\multicolumn{7}{c}{} \\ [0.5ex]
 \hline 
  &FSC &FIC&FFSC& FFIC&SFSC&SFIC\\ 
 \hline
$d_{E,\min}^2$ &\checkmark&&\checkmark&&\checkmark&\\
$\tau_E$&\checkmark&&\checkmark&&\checkmark&\\
 $d_{P,\min}^2$&&\checkmark&&\checkmark&&\checkmark\\
$\tau_P$&&\checkmark&&\checkmark&&\checkmark\\
$L$&&\checkmark&&\checkmark&&\checkmark\\
$N_d$ &\checkmark&\checkmark&\checkmark&\checkmark&\checkmark&\checkmark\\
 Bit-labeling&\checkmark&\checkmark&\checkmark&\checkmark&\checkmark&\checkmark\\
 \hline 
\end{tabular}}}
\label{table:kpis} 
\end{table}
\begin{table*}[!t]
\caption{{\color{black}Best known multidimensional constellations for different channel scenarios}}
\centering 
{\color{black}
\scalebox{0.8}{
\begin{tabular}{ccc}
\multicolumn{3}{c}{} \\ [0.7ex]
 \hline 
  Channel Scenario&Best 4-point Constellation(s)&Best 16-point Constellation(s)\\ 
  \hline
 FSC (SER) &4-Beko&16-Beko,  16-Bao, T16QAM\\
FSC (BER)&4-LDS&16-Bao, T16QAM\\
 FIC (SER)&4-Bao, 4-LDS&16-Bao, T16QAM\\
FIC (BER)&4-Bao, 4-LDS&16-Bao, T16QAM\\
High-rate FFSC (FER)&4-LDS&16-Bao, T16QAM\\
High-rate FFIC (FER) &4-LDS&16-Bao, T16QAM\\
 High-rate SFSC (FER)&4-Beko, 4-Bao, 4-LDS&16-Bao, T16QAM\\
 High-rate SFIC (FER)&4-Bao, 4-LDS&16-Bao, T16QAM\\
Low-rate FFSC (FER) &4-LDS&16HQAM\\
 Low-rate FFIC (FER)&4-Bao, 4-LDS&16HQAM\\
  Low-rate SFSC (FER) &4-LDS, 4-Bao, 4LQAM, 4CQAM, T4QAM&16HQAM, T16QAM, 16LQAM, 16-Bao\\
 Low-rate SFIC (FER)&4-Bao, 4-LDS&T16QAM\\
 \hline 
\end{tabular}}}
\label{table:results} 
\end{table*}
In Table \ref{table:kpis}, we provide a summary of KPIs that should be considered in the design process of multidimensional constellations for uplink SCMA systems over different scenarios.

Based on the results provided in Section \ref{subsec:uc}--\ref{subsec:lowratetc}, we summarize the best 4-point and 16-point constellation(s) for different scenarios in Table \ref{table:results}. Note that the objective of the current work is not to determine what the best constellation is, rather to provide guidance of designing new efficient constellations for uplink SCMA systems.
}

\section{Discussions and Future Directions}\label{sec:conlusion}
In this paper, we provided a comparative study on some of the most important existing $M$-point $2d_v$-dimensional constellations for SCMA systems under various channel conditions. Since SCMA has been proposed to support uplink machine-type communication services, we have only focused on uplink SCMA systems. We began our discussion with identifying the KPIs of constellations that are of significance for different channel scenarios. We then presented different existing constellations with their design criteria, and evaluated the performance of an important subset of those constellations for uncoded, high-rate and low-rate LTE turbo-coded SCMA systems under various channel scenarios. All turbo-coded comparisons were performed for BICM, with a concatenated detection and decoding scheme.

 {\color{black} This paper sheds some light in designing multidimensional constellations for SCMA systems over a variety of channel conditions.} When designing an $M$-point $2d_v$-dimensional constellation for uplink uncoded SCMA systems over FSC and high-rate turbo-coded systems over FFSC and SFSC, it is imperative to consider {\color{black}maximizing $N_d$} and $d_{E,\min}$, and minimizing $\tau_E$. In systems with a low rate code over FFSC and SFSC, the effect of {\color{black}$N_d$} is of less concern. On the other hand, in the design process of an $M$-point $2d_v$-dimensional constellations for uplink SCMA systems over FIC, FFIC, and SFIC, it is desired to {\color{black}maximize $N_d$}, minimize $L$, maximize $d_{P,\min}$, and minimize $\tau_P$. Furthermore, the BER performance of the uncoded systems, and the performance of the coded systems are tied to their bit-labeling. The performance of the systems also highly depends on the behavior of the multi-user detector at different SNRs. 

{\color{black}In summary, the behavior of multidimensional constellations changes with the dynamic of the channel. In other words, different constellations perform differently over various channel scenarios. As such, in order to maximize the gain in a certain channel scenario, a proper multidimensional constellation can be designed using the specific KPIs for that scenario.}

A number of possible directions in the line of multidimensional constellations for SCMA systems is as follows:
\begin{itemize}
  \item In this paper, it is assumed that CSI is not available at the transmitter. {\color{black}In some scenarios it is essential to obtain CSI at the transmitter. The impact of total CSI or partial CSI at the transmitter on the behaviour of multidimensional constellations can be investigated. {\color{black}For example, it is likely that the user-specific 2-dimensional rotations pointed out in Section \ref{sec:constdesign} are useful when CSI is available at the transmitter.}}
  \item {\color{black}It is assumed that each user and the receiver are accompanied with a single transmit and a single receive antenna, i.e., SISO-SCMA. Alternatively, employing multiple transmit and receive (MIMO) antennas  can improve the system performance by exploiting the spatial domain. As such, the effect of MIMO antennas, i.e., MIMO-SCMA (e.g., \cite{han16mimoscma,Abdessamad16mimoscma}) on the performance multidimensional constellations can be studied.}
  \item {\color{black}For the coded scenarios, only the LTE turbo code has been {\color{black}considered here}. It can be verified whether the KPIs provided in designing multidimensional constellations depend on the channel coding techniques or not. Therefore, the behavior of multidimensional constellations in conjunction with other channel coding techniques such as LDPC codes (\cite{xiao15,hao17polar}) or polar codes (e.g., \cite{vameghestahbanati2017polar,Dai16_polar,jing17polar,pan18polar}) can be further examined.}
  \item All the comparisons are made for BICM. {\color{black}It is shown in the literature  that for some applications (e.g., \cite{Seidl13}) multilevel coding (MLC) \cite{MLC_Imai} exhibits some gain over its BICM counterpart.} Studying the KPIs for MLC is a possible research direction.
  \item It is assumed that the detection and decoding are performed separately in a non-iterative manner. {\color{black}It is shown in e.g., \cite{wu15} that an iterative multiuser receiver improves the SCMA system performance.} The effect of iterative decoding and detection{\color{black}, i.e., BICM with iterative decoding (BICM-ID)}, (e.g., \cite{wu15,Bao18,Bao17Icc}) can be investigated.
  \item {\color{black}One of the main challenges in SCMA systems is the receiver complexity that inherits from the detection of all active users \cite{bayesteh15}. In this paper, the widely-used non-binary MPA is used as the multi-user detection.} The impact of the choice of constellation when other lower complexity detectors (e.g., \cite{wei2016low,meSD1CL17,du16detection2,Du16detection,du16,wang16detection,Belfiore17,Meng17detection,dai17detection,zhang18,Jia18,Li18SD})
are used, should be further studied. 
\item All the comparisons in this paper have performed for a regular user-to-RE mapping matrix. {\color{black}It was also shown in \cite{Shental17}, that a regular user-to-RE allocation is advantageous.} {\color{black}Nevertheless, in a regular SCMA structure, it is not possible to serve different users with various requirements. Thus, studying the effect of an irregular SCMA structure, whereby users are not forced to occupy only a fixed number of REs, e.g., \cite{zhang15irregular,Yu16irregular,jiangldsm17},} is a possible research direction.
   \item {\color{black}{\color{black}Mathematical analysis of} the error performance of constellations provides more indication on their performance in different SNR regions.} In \cite{Cagri16CL,Cagri18,Bao16pep,Bao17pep,cai17SER,Tian18}, the performance of some of the constellations has been analyzed. The performance analysis of different constellations under different scenarios is a possible research direction.
  \item {\color{black} One of the problems with multi-carrier systems is their high peak-to-power-average-ratio (PAPR). Since SCMA can be regarded as a type of multi-carrier NOMA \cite{Ding17survey}, the behavior of different multidimensional constellations from PAPR perspective can be studied along with PAPR reduction techniques, e.g., \cite{yang18papr}.}
  \item {\color{black}The KPIs of designing multidimensional constellations for other channel models that are out of the scope of the current work can be explored.}
  \item {\color{black}Obtaining the relationship between the KPIs that we have obtained for SCMA systems with the KPIs of single-user systems over different scenarios can be considered.} 
  \item The KPIs that are given for each specific scenario can be considered in the design process of new multidimensional constellations over each scenario.
\end{itemize}
\section{Acknowledgement}
{\color{black}The authors would like to thank Dr. Rui Dinis from the Faculdade de Ciências e Tecnologia,
Universidade Nova de Lisboa, Lisbon, Portugal, {\color{black}and the anonymous reviewers, for their} valuable feedback.}
\appendices
\section{{\color{black}Effects of user-specific rotations}}\label{app:userspecific}
The optimization of 2-dimensional user-specific rotations is of importance only in downlink scenarios \cite{taherzadeh14}, 
because in uplink SCMA systems the users experience different fading channels, {\color{black}so the benefits of user-specific rotations are neglected.} 
For illustrative purposes, consider a downlink and an uplink scenario with two users depicted in Fig.~\ref{fig:downlink} and Fig.~\ref{fig:uplink}. \begin{figure}[t]
\centering     
\subfigure[]{\label{fig:downlink}\includegraphics[width= 0.225\textwidth]{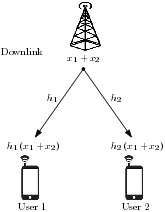}}
\subfigure[]{\label{fig:uplink}\includegraphics[width= 0.2\textwidth]{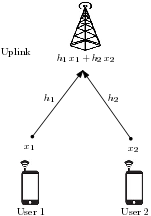}}
\caption{The illustration of the importance of user-specific rotations in (a) a downlink, and (b) an uplink scenario.}
\end{figure}
Let $x_1$ and $x_2$ denote the symbols corresponding to User 1 and User 2, and $h_1$ and $h_2$ represent the fading channel coefficients of User 1 and User 2, respectively. As we see in Fig.~\ref{fig:downlink}, in a downlink scenario each user receives a combination of both symbols that are rotated and scaled by the same channel coefficient, i.e., $h_1\left(x_1+x_2\right)$ at User 1, or $h_2\left(x_1+x_2\right)$ at User 2. {\color{black}Careful selection of the user-specific rotations in $x_1$ and $x_2$ facilitates recovering} the symbol of each user {\color{black}from the sum, $x_1+x_2$}. In contrast, as we see in Fig.~\ref{fig:uplink}, in an uplink scenario when both users experience different fading channels, each symbol corresponding to each user is rotated and scaled by a different random value, i.e., $x_1\:h_1$ and $x_2\:h_2$. As such, any benefit of a 2-dimensional rotation on each symbol {\color{black}prior to transmission will be negated by the random rotation caused by} random fading channel coefficients. Restated, a 2-dimensional rotation on each RE has no impact in an uplink scenario. 
\bibliographystyle{IEEEtran}
\bibliography{IEEEabrv,mybib2}
\end{document}